\documentclass[reprint,
nofootinbib,
amsmath,
amssymb,
aps,
pra]{revtex4-2}

\usepackage{graphicx}
\usepackage{dcolumn}
\usepackage{bm}
\usepackage{makecell}
\usepackage{cancel}
\usepackage{hyperref}
\usepackage[dvipsnames]{xcolor} 
\usepackage{comment}
\usepackage[capitalise]{cleveref}

\begin{document}

\preprint{APS/123-QED}

\title{Simulating imperfect quantum optical circuits using unsymmetrized bases}

\author{John Steinmetz}
\email{jcsteinmetz@gmail.com}
\author{Maike Ostmann}
\author{Alex Neville}
\author{Brendan Pankovich}
\author{Adel Sohbi}
\affiliation{
ORCA Computing
}

\date{\today}

\begin{abstract}
Fault-tolerant photonic quantum computing requires the generation of large entangled resource states. The required size of these states makes it challenging to simulate the effects of errors such as loss and partial distinguishability. For an interferometer with $N$ partially distinguishable input photons and $M$ spatial modes, the Fock basis can have up to ${N+NM-1\choose N}$ elements. We show that it is possible to use a much smaller unsymmetrized basis with size $M^N$ without discarding any information. This enables simulations of the joint effect of loss and partial distinguishability on larger states than is otherwise possible. We demonstrate the technique by providing the first-ever simulations of the generation of imperfect qubits encoded using quantum parity codes, including an example where the Hilbert space is over $60$ orders of magnitude smaller than the $N$-photon Fock space. As part of the analysis, we derive the loss mechanism for partially distinguishable photons.
\end{abstract}

\maketitle

\section{Introduction}
A linear optical quantum computer is made up of circuits that generate and manipulate multiphoton entangled resource states~\cite{Knill2001,Kok2007}. These circuits are composed of linear optical components such as beam splitters and phase shifters, which in practice can each act imperfectly on the quantum state. It is important to understand how different imperfections in these circuit components affect the quantum state in order to predict the output of the circuit and determine the best way to correct any errors that occur. Quantum circuit simulators can be used to classically model these imperfect circuits, so they are a valuable tool to guide the design of a physical device. Several simulators intended specifically for photonic circuits have been released in recent years, coinciding with the rapid growth of photonic quantum computing as a field~\cite{BosonicQiskit, StrawberryFields, Perceval, SOQCS, PhotoniQLAB, Piquasso}.

Partial distinguishability is an important source of error to accurately model in optical circuits due to the practical difficulty of producing identical photons. If the photons in a circuit have different properties, then they do not perfectly interfere. The output of a boson sampling experiment with partial distinguishability can be modelled using matrix permanents~\cite{Tichy2017,Moylett2018,Renema2018}. There have also been several publications in recent years that incorporate some version of partial distinguishability in a linear optical circuit simulation~\cite{PhotoniQLAB, Osca2023, Saied2024}. When these simulations use density matrices (or state vectors), the partial distinguishability is modelled by adding extra modes to the system corresponding to the additional properties of the photons. This increases the space size, so there is a significant restriction on the size of circuits that can be simulated. The larger space size also makes it more difficult to model other imperfections such as loss, so the problem of combining loss with partial distinguishability is largely unexplored.

An alternative to density matrix simulations is to use matrix permanents, which are frequently used in the context of boson sampling~\cite{Aaronson2013,Clifford2017}. For a system of indistinguishable photons, permanents can be used to reconstruct the full density matrix at the output of a circuit. However, when the photons are partially distinguishable, this method only yields output probabilities, not amplitudes; reconstructing the full state can be done with tomographic reconstruction~\cite{Shaw2023}, which has a higher computational cost. It is also more difficult to deal with processes like loss that yield mixed states. These can be modelled in a way that keeps the state pure, like adding environment modes and expanding the working space. In certain special cases, loss can be commuted to the beginning of the circuit, which allows each component of the mixed state to be treated separately. Despite these limitations, the process of computing matrix permanents is highly optimized, so this is often the method of choice for simulating circuits.

In this paper, we propose a new set of density matrix simulation techniques that can be used to model systems that include both partially distinguishable photons and non-uniform loss in a way that scales more favourably with the circuit size than previously published methods. This is done by working in an unsymmetrized space instead of the usual Fock space, delaying the photons from interfering until the end of the circuit. We propose a specific unsymmetrized basis for a system of partially distinguishable photons such that the dimension of the Hilbert space scales better than Fock space. The new framework naturally handles mixed states since states are still represented as density matrices, so we can implement non-uniform loss without introducing any extra modes. In order to model loss and distinguishability together, we derive generalized Kraus operators and show how to use them in the unsymmetrized space. We demonstrate the new method by obtaining complete density matrices for some example circuits in the presence of arbitrary loss and partial distinguishability, focusing on generators of entangled resource states that are included in existing proposals for fault-tolerant devices.

In Section~\ref{sec:partial}, we review the theory of partially distinguishable photons in first and second quantization. In Section~\ref{sec:unsym}, we propose an unsymmetrized basis and show how it can result in a smaller space size. In Section~\ref{sec:resolve}, we demonstrate how to obtain physically meaningful results from the unsymmetrized density matrix. In Section~\ref{sec:loss}, we show how to combine loss with partial distinguishability and integrate it into the proposed framework. In Section~\ref{sec:examples}, we demonstrate our new simulation techniques on some example circuits, including generators of encoded qubits. In Section~\ref{sec:conclusion}, we conclude and discuss the merits of our proposal relative to other simulation techniques.

\section{Partially distinguishable photons}
\label{sec:partial}
\subsection{Internal states}
Each particle in a circuit, in addition to its spatial location, has other properties that could potentially distinguish it from other particles. For example, each particle might have a different spectral profile, time delay or polarization. We can express each particle's state as the tensor product of an \emph{external state} $|m\rangle$, which refers to its location in the circuit, and an \emph{internal state} $|\phi\rangle$, which contains all its other properties~\cite{Tichy2014}. Note that at this stage we are using the word ``particle" instead of the word ``photon", because until the state is symmetrized, the analysis applies to both bosons and fermions. The single-photon Hilbert space is
\begin{equation}
    \mathcal{H}=\mathcal{H}_\text{ext}\otimes\mathcal{H}_\text{int},
\end{equation}
where the external and internal spaces are spanned by the bases
\begin{equation}
    \mathcal{B}_\text{ext}=\{|m\rangle\},\quad \mathcal{B}_\text{int}=\{|\phi\rangle\},
\end{equation}
where $m\in\{1,\dots,M\}$.

While the external modes are all mutually orthogonal, this is not necessarily the case for the internal modes. Each pair of internal modes has an overlap $\langle\phi_i|\phi_j\rangle\equiv S_{ij}$, which can be considered elements of a distinguishability matrix $S$~\cite{Tichy2015}. The square of these overlaps $|S_{ij}|^2$ is the pairwise visibility, which in a photonic system quantifies the size of the fringes that appear when two photons are interfered at a beam splitter, as in a Hong-Ou-Mandel experiment~\cite{HOM1987,Branczyk2017}. A visibility of $|S_{ij}|^2=0$ between two photons means they are perfectly distinguishable and will not interfere. If all the photons in a system are distinguishable, then the system will behave classically. Ideally, all pairs of photons have perfect visibility $|S_{ij}|^2=1$, which means they are identical and will interfere completely. 

A system becomes more difficult to model when there are partially distinguishable photons, which can have a range of visibilities with $0<|S_{ij}|^2<1$. The transition between distinguishable and indistinguishable particles is not a linear interpolation~\cite{Tichy2015}.

The usual method of dealing with partial distinguishability is to transform the internal states into an orthogonal basis 
\begin{equation}\label{eq:ortho_int_basis}
    \tilde{\mathcal{B}}_{\text{int}}=\{|\tilde{\phi}\rangle\}
\end{equation}
using a method such as the Gram-Schmidt process~\cite{NielsenBook, Tichy2014}. Although this can be a useful technique, we will not use it here. \cref{app:gs} covers orthogonalization and our reasons for avoiding it in more detail. By working in an unsymmetrized space, we avoid having to orthogonalize the internal states, modelling the system using only the pairwise overlaps $S_{ij}$.

\subsection{Effective distinguishability}
For a system of $N$ particles, the Hilbert space is a tensor product of single-particle spaces. The tensor product can be rearranged to write it as a product of external and internal states
\begin{equation}
    \mathcal{H}^{\otimes N}=\mathcal{H}^{\otimes N}_{\text{ext}} \otimes \mathcal{H}^{\otimes N}_{\text{int}}
    \label{eq:HN}
\end{equation}
spanned by the basis
\begin{equation}
    \mathcal{B}^{\otimes N}=\{|\bm{m},\bm{\phi}\rangle : |\bm{m}\rangle\in\mathcal{B}^{\otimes N}_{\text{ext}}, |\bm{\phi}\rangle\in\mathcal{B}^{\otimes N}_{\text{int}}\},
    \label{eq:basis-general}
\end{equation}
where $|\bm{m},\bm{\phi}\rangle=|\bm{m}\rangle\otimes|\bm{\phi}\rangle$. Vector notation will be used to represent tensor products of kets, so $|\bm{m}\rangle=\bigotimes_{i=1}^N |m_i\rangle$ and $|\bm{\phi}\rangle=\bigotimes_{i=1}^N |\phi_i\rangle$, where $|m_i\rangle$ and $|\phi_i\rangle$ are the $i$th particle's external state and internal state, respectively. We will often make the assumption that there are $N$ possible internal modes that can be occupied, one for each photon. If that is the case, then the size of this basis is 
\begin{equation}
    \text{dim}(\mathcal{H}^{\otimes N})=(NM)^N,
    \label{eq:size-unsym}
\end{equation}
since we are selecting from $NM$ total modes for each of the $N$ particles.

The elements of the $N$-particle basis are not symmetric or anti-symmetric. When particles are exchanged by permuting their positions in the tensor product, there are certain basis states that are mapped to a different basis state. For example, consider the two states
\begin{equation}
    |12\rangle|\phi_1\phi_2\rangle \neq |21\rangle|\phi_2\phi_1\rangle,
    \label{eq:1221}
\end{equation}
where exchanging the two particles yields a different state. Although the particles on either side of this expression have the same external and internal state, they appear in different positions in the tensor product. In other words, this basis state is not invariant under particle exchange.

Bosonic systems are symmetric under particle exchange, so the two sides of this expression should be mapped to the same physical state
\begin{equation}
    a_1^\dagger(\phi_1)a_2^\dagger(\phi_2)|\text{vac}\rangle,
\end{equation}
where $a_i^\dagger(\phi_j)$ creates a boson in external mode $i$ and internal mode $\phi_j$. Unlike this physical state, which is symmetric under particle exchange, the two states in \cref{eq:1221} have an extra degree of freedom that distinguishes the two particle orderings. So in the $N$-photon space we have described, the particles are not yet being treated as bosons (or fermions).

The position of a particle in the tensor product space is called its \emph{label}, $\ell_i$. On the left hand side of \cref{eq:1221}, particle $\ell_1$ is in the state $|1,\phi_1\rangle$ and particle $\ell_2$ is in the state $|2,\phi_2\rangle$. On the right hand side, these labels are swapped. In the literature, external and internal modes are sometimes referred to as ``system" and ``label" modes~\cite{Turner2016,Stanisic2018,JonesThesis}, but we will use the term ``label" exclusively to refer to particle ordering. In an example like \cref{eq:1221}, the presence of the label degree of freedom makes it possible to know which particle has which external and internal state. We will refer to this kind of distinguishability, where particles can be told apart using their label information, as \emph{effective distinguishability}~\cite{BrunnerMasters}. In order for states to interfere as they should in a bosonic system, the effective distinguishability needs to be eliminated by symmetrizing the space.

\subsection{First quantization}
A bosonic system occupies the part of the $N$-photon Hilbert space that is symmetric in the label degree of freedom,
\begin{equation}
    \mathcal{H}_S=\mathcal{S}(\mathcal{H}^{\otimes N}).
\end{equation}
In this subspace, different particles cannot be told apart by their label, although it is still possible for them to be distinguishable through their internal states.

The symmetrization operator that projects onto the symmetric subspace is defined by the transformation
\begin{equation}
    \mathcal{S}|\bm{m},\bm{\phi}\rangle=\frac{1}{\sqrt{|S_N||I_{\bm{m},\bm{\phi}}|}}\sum_{\pi\in S_N} \pi|\bm{m},\bm{\phi}\rangle
    \label{eq:S}
\end{equation}
where $\pi$ are particle exchange operators drawn from the symmetric group $S_N$. Each $\pi$ permutes the particles' labels in a different way, and the sum includes all such permutations. It is possible, depending on the state, that there is a subset $I_{\bm{m},\bm{\phi}}$ of label permutations under which the state is invariant. This can happen if there are multiple particles in the same combination of external and internal modes, in which case permuting the labels of those particles will not alter the state. For example, permuting the two particles in the state $|11,\phi_1\phi_1\rangle$ has no effect.

The set $S_N$ and subset $I_{\bm{m},\bm{\phi}}$ have sizes
\begin{equation}
    \begin{split}
        |S_N|&=N! \\
        |I_{\bm{m},\bm{\phi}}|&=\prod_{m,\phi}N_{m,\phi}!,
    \end{split}
    \label{eq:norm}
\end{equation}
where $N_{m,\phi}$ is the number of photons in external mode $m$ and internal mode $\phi$. The presence of these constants in \cref{eq:S} keeps the state normalized under symmetrization.

Applying this operator to every basis state in \cref{eq:basis-general} gives the symmetrized basis
\begin{equation}
    \mathcal{B}_{1Q}=\{\mathcal{S}|\bm{m},\bm{\phi}\rangle:|\bm{m}\rangle\in\mathcal{B}_\text{ext}^{\otimes N},|\bm{\phi}\rangle\in\mathcal{B}_\text{int}^{\otimes N}\},
    \label{eq:basis-1q}
\end{equation}
where duplicates are removed. This formalism is \emph{first quantization} (1Q)~\cite{Beggi2018,BrunnerMasters}, where each entry in a ket is the state of a specific photon. The particles are no longer effectively distinguishable in this basis since the basis elements are all symmetric under particle exchange. However, the particles are still partially distinguishable as a result of their overlapping internal states, so the basis elements are not mutually orthogonal.

As an example of first-quantized basis states, we can symmetrize both sides of \cref{eq:1221}. Both sides give the same state,
\begin{equation}
    \begin{split}
        \mathcal{S}(|12\rangle|\phi_1\phi_2\rangle)&=\frac{|12\rangle|\phi_1\phi_2\rangle+|21\rangle|\phi_2\phi_1\rangle}{\sqrt{2}} \\
        \mathcal{S}(|21\rangle|\phi_2\phi_1\rangle)&=\frac{|12\rangle|\phi_1\phi_2\rangle+|21\rangle|\phi_2\phi_1\rangle}{\sqrt{2}}.
    \end{split}
    \label{eq:1221-symmetrized}
\end{equation}
So by applying the symmetrization operator, all the elements of $\mathcal{B}^{\otimes N}$ that correspond to the same physical state get mapped to the same element of the symmetric basis $\mathcal{B}_{1Q}$. This eliminates the label degree of freedom and causes the particles to interfere in the correct way for bosons.

If the system were fermionic, we would need to project onto the set of anti-symmetric basis states instead. We do not cover fermions here, but similar analyses in the context of partial distinguishability have been done elsewhere~\cite{Beggi2018,BrunnerMasters}.

\subsection{Second quantization}
The symmetrized basis states in \cref{eq:basis-1q} can be written more concisely by moving to \emph{second quantization} (2Q), where instead of listing the locations of each photon in the circuit, we list the number of photons in each mode. For example, the state from the previous section can be rewritten
\begin{equation}
\frac{|12\rangle|\phi_1\phi_2\rangle+|21\rangle|\phi_2\phi_1\rangle}{\sqrt{2}}\leftrightarrow|10\ 01\rangle\rangle,
\end{equation}
where the double ket is used to indicate that its contents are occupation numbers. Each entry is the number of photons in the mode combination $(m,\phi)$, and spaces are used to separate different external modes. In this example, there is one photon in the first external mode with internal mode $\phi_1$, and another photon in the second external mode with internal mode $\phi_2$. The basis of second-quantized states, also called the Fock basis, is
\begin{equation}
    \mathcal{B}_{2Q}=\{|\bm{n}\rangle\rangle:\sum_i n_i = N\}.
    \label{eq:basis-2q}
\end{equation}
Note that we will always use $m$ to refer to external modes and $n$ to refer to occupation numbers. 

There is a one-to-one correspondence between first- and second-quantized basis states,
\begin{equation}
     \mathcal{S}|\bm{m},\bm{\phi}\rangle\leftrightarrow|\bm{n}\rangle\rangle,
     \label{eq:conversion}
\end{equation}
so the Fock basis also spans $\mathcal{H}_S$. Just like with first quantization, the elements of this basis are not mutually orthogonal due to the overlapping internal modes, so $\langle\langle\bm{n}_i|\bm{n}_j\rangle\rangle\neq\delta_{ij}$.

The number of Fock basis elements is
\begin{equation}
     \text{dim}(\mathcal{H}_S) = {N+NM-1\choose N},
    \label{eq:size-sym}
\end{equation}
which is the number of ways to distribute $N$ photons into $NM$ modes. Although this is still impractical, it is smaller than that of the unsymmetrized space $\mathcal{H}^{\otimes N}$ \cref{eq:size-unsym}, which contains symmetric states, anti-symmetric states, and states with mixed symmetry. In order for the unsymmetrized space to be useful in practice, we need to reduce its dimension to a lower number than \cref{eq:size-sym}.

\section{Unsymmetrized bases}
\label{sec:unsym}

\subsection{Delaying symmetrization}
In a density matrix simulation, the quantum state is usually written in a symmetrized basis like the Fock basis~\cite{Piquasso, BosonicQiskit, SOQCS}. In that case, the symmetrization procedure occurs implicitly before the simulation begins. Throughout the simulation, the state and the circuit operations are stored as matrices in the Fock space $\mathcal{H}_S$. In order to avoid using this space, we propose to apply the circuit operation before symmetrizing the state, allowing the state and circuit operations to be stored in an unsymmetrized basis. Although the full unsymmetrized space $\mathcal{H}^{\otimes N}$ is not very useful since it is larger than the symmetrized space $\mathcal{H}_S$, it is possible to find unsymmetrized bases that are much smaller and provide a significant advantage.

The circuit operation, which has the operator-sum representation~\cite{NielsenBook}
\begin{equation}
    \Lambda(\rho)=\sum_i E_i\rho E_i^\dagger,
    \label{eq:op-sum}
\end{equation}
can be performed before or after symmetrization~\cite{Beggi2018,BrunnerMasters}. Each operation element $E_i$ should be invariant under photon label exchange so that it treats a particle the same way regardless of its position in the tensor product space. This means that for any particle exchange operator $\pi$,
\begin{equation}
    [E_i,\pi] = 0.
\end{equation}
The symmetrization operator $\mathcal{S}$, defined in \cref{eq:S}, is a sum of exchange operators, so
\begin{equation}
    [E_i,S] = 0,
\end{equation}
which leads to the result
\begin{equation}
    \Lambda(\mathcal{S}\rho\mathcal{S}^\dagger)=\mathcal{S}\Lambda(\rho)\mathcal{S}^\dagger.
\end{equation}
This means it is possible to change the order of operations and apply the circuit operation to an unsymmetrized state then symmetrize the output. This turns out to be a powerful result since, as we will discuss in the rest of this section, working with unsymmetrized states allows the space size to be reduced to a point where it is smaller than the symmetrized basis $\mathcal{H}_S$.

\subsection{Mode assignment lists}

The unsymmetrized basis $\mathcal{B}^{\otimes N}$, defined in \cref{eq:basis-general}, contains elements that differ only by a label permutation. When the space is symmetrized, these label-permuted elements will be mapped to the same symmetric state. For example, the two states we looked at in \cref{eq:1221} have different labellings but represent the same physical state, so they get mapped to the same basis state under symmetrization. Even though the label degree of freedom is not physically meaningful, it still contributes to the size of the unsymmetrized space.

To reduce the space size, we can use a photon's label as a placeholder for one of its other properties as long as that property is constant through the entire circuit and does not affect the action of the circuit components. For example, consider a system where $\Lambda$ only acts on the external modes, moving photons around the circuit without affecting their internal state. In that case, if a photon with label $\ell$ has a certain internal state $\phi$, then it will have that same internal state at every point in the circuit, so there is no reason to store both pieces of information.

\begin{figure}
    \centering
    \includegraphics[trim=0 0 0 0, clip, width=0.9\columnwidth]{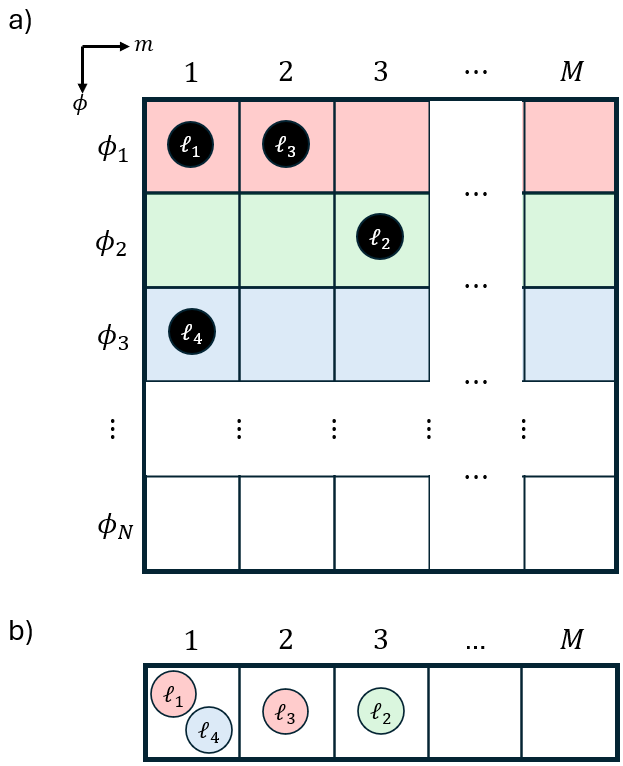}
    \caption{An illustration of a) the full $N$-particle unsymmetrized space $\mathcal{H}^{\otimes N}$ and b) the space of external mode assignment lists $\mathcal{H}_\text{MAL}$, which is condensed but equivalent under symmetrization. Photons are represented as discs containing a label $\ell_i$ indicating the photon's position in the tensor product. In the unsymmetrized state $|1321\rangle|\phi_1\phi_2\phi_1\phi_3\rangle$, each photon occupies an external mode and an internal mode. In the corresponding mode assignment list $|1321)$, each photon's label is used as a stand-in for its internal state. We know that the photon with a certain label has a certain internal state, so the two pieces of information can be condensed into a single degree of freedom. Even though the two photons $\ell_1$ and $\ell_3$ have the same internal state, they are effectively distinguishable because of their distinct labels so they will not interfere until the state is symmetrized.}
    \label{fig:mal}
\end{figure}

Instead, we can associate each position in the ket with a specific internal state. For example, consider $|\bm{m},\bm{\phi}\rangle=|1321\rangle|\phi_1\phi_2\phi_1\phi_3\rangle$, a particular state of four photons with various internal states in four external modes. We can make the association $(\ell_1,\ell_2,\ell_3,\ell_4)\leftrightarrow(\phi_1,\phi_2,\phi_1,\phi_3)$ and write the system state as the external \emph{mode assignment list} (MAL)~\cite{Tichy2012,Englbrecht2023}
\begin{equation}
    |\bm{m})=|\underset{\ell_1}{1},\underset{\ell_2}{3},\underset{\ell_3}{2},\underset{\ell_4}{1}),
\end{equation}
where we choose the labels to always be ordered, so they will usually not be written underneath the list. The internal states do not need to be stored, since each label corresponds to a specific internal state. In other words, we know a photon's internal state by its position in the mode assignment list. The notation $|.)$ will be used for mode assignment lists, where the order of labels is fixed, to distinguish them from standard kets where the labels can be chosen to be arranged in any order. \cref{fig:mal} contains an illustration of this example.

Every mode assignment list implicitly contains the same ``un-permuted" set of internal states, $|\bm{\phi}^{(0)}\rangle$. In the previous example, we chose the order $(\phi_1,\phi_2,\phi_1,\phi_3)$, but there is freedom to use any permutation of $|\bm{m},\bm{\phi}\rangle$. In the literature, mode assignment lists are often written with the external modes in ascending order, but we will always write them in the order determined by $|\bm{\phi}^{(0)}\rangle$.

Since the space of mode assignment lists is unsymmetrized, the photons are effectively distinguishable by their labels. This is true regardless of their internal states. No matter whether the photons are perfectly distinguishable, partially distinguishable, or identical, we can always evolve the state using distinguishable-particle operators, tracking where each specific photon ends up in the circuit. Afterwards, we can implement the interference by symmetrizing the space, a process that is covered in \cref{sec:resolve}. This effective distinguishability makes mode assignment lists a natural way of handling partially distinguishable photons\footnote{It is also possible to use the label information as a placeholder for external modes instead of internal modes. For the same example state, we can associate each label with an external mode, $(\ell_1,\ell_2,\ell_3,\ell_4)\leftrightarrow(1,3,2,1)$, and write the state as an \emph{internal mode assignment list},
\begin{equation}
    |\bm{\phi})=|\underset{\ell_1}{\phi_1},\underset{\ell_2}{\phi_2},\underset{\ell_3}{\phi_1},\underset{\ell_4}{\phi_3}).
\end{equation}
Each photon's position in this list of internal modes tells us its external mode. This version is less useful for modelling interferometers; an analogous application would be a circuit where components only affect the photons' internal modes without affecting their spatial location. For now, we will mainly focus on external mode assignment lists because of their application to interferometers. Unless specified, any mention of mode assignment lists will refer to the external type.}.

\begin{figure}
    \centering
    \includegraphics[trim=10 10 0 0, clip, width=\columnwidth]{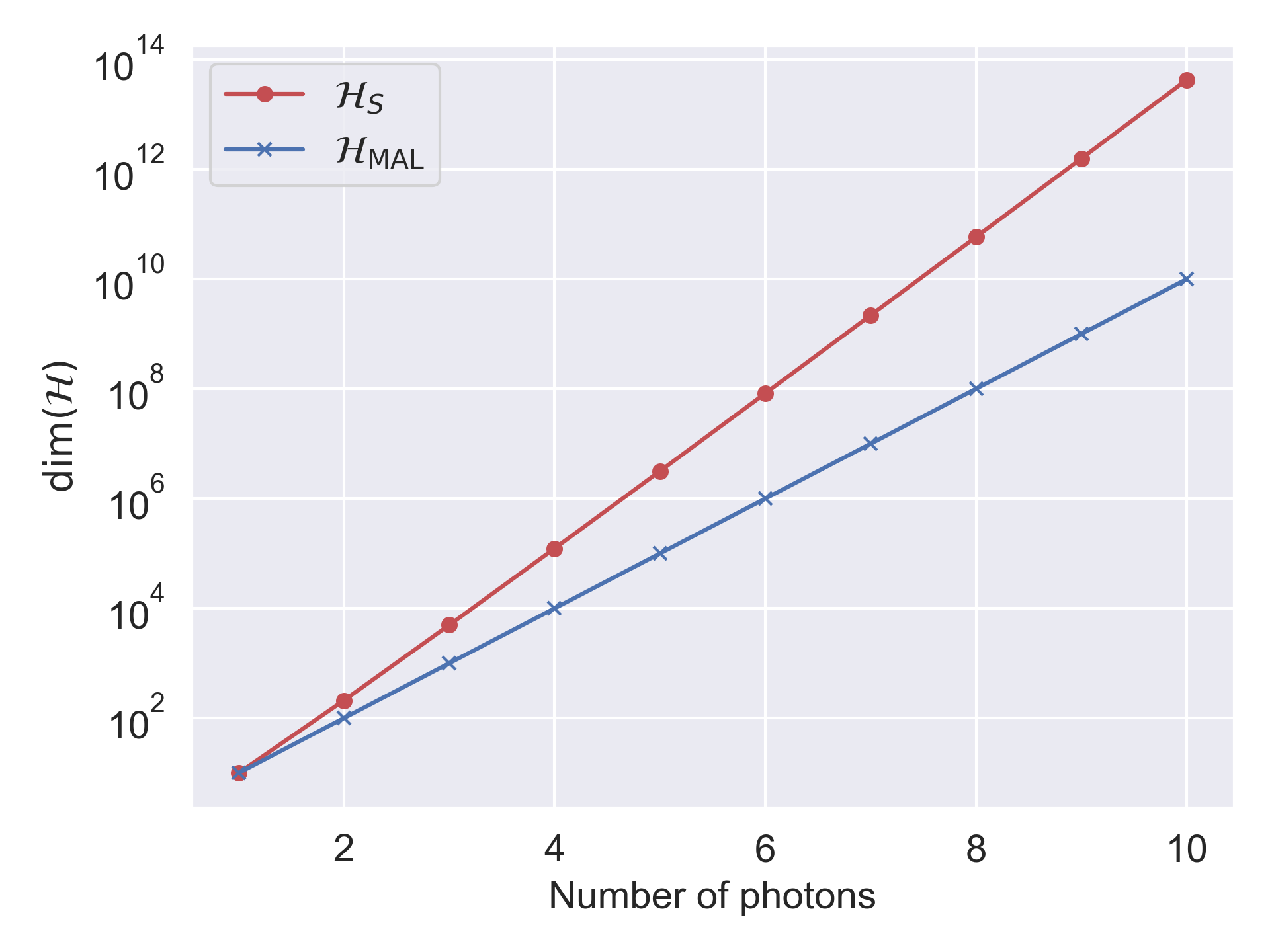}
    \caption{The size of the symmetrized Hilbert space \cref{eq:size-sym} compared to the size of the Hilbert space for mode assignment lists \cref{eq:size-mal}. The number of external modes is held constant at $M=10$ and the scaling is shown as more photons are added to the circuit. The mode assignment list space scales more favourably than the symmetrized space. When the system contains ten photons, there is an advantage of almost four orders of magnitude.}
    \label{fig:malscaling}
\end{figure}

The space of mode assignment lists is the same as the the $N$-particle external Hilbert space,
\begin{equation}
    \mathcal{H}_{\text{MAL}}=\mathcal{H}_\text{ext}^{\otimes N}
\end{equation}
which is spanned by the basis
\begin{equation}
    \mathcal{B}_\text{MAL}=\{|\bm{m}): |\bm{m})\in\mathcal{B}_\text{ext}^{\otimes N}\},
\end{equation}
where each basis element
\begin{equation}
    |\bm{m})=|\bm{m}\rangle\otimes|\bm{\phi}^{(0)}\rangle
\end{equation}
implicitly contains the same chosen ``un-permuted" list of internal states $|\bm{\phi}^{(0)}\rangle$. Since the internal modes do not contribute to the space size, the number of basis elements is only
\begin{equation}
    \text{dim}(\mathcal{H}_{\text{MAL}}) = M^N,
    \label{eq:size-mal}
\end{equation}
where each of the $N$ photons can be located in any of the $M$ external modes. This is smaller than the size of the symmetrized space $\mathcal{H}_S$. \cref{fig:malscaling} shows an example of $\text{dim}(\mathcal{H}_{\text{MAL}})$ scaling more favourably than $\mathcal{H}_S$ for $M=10$ external modes. \cref{tab:bases} contains a summary of all the bases we have covered so far and their properties.

\begin{table}
    \centering
    \begin{tabular}{|c|c|c|c|c|}
        \hline
         Basis name & Symbol & Symm. & Size & Example \\
         \hline
         \makecell{$N$-particle \\ basis} & $\mathcal{B}^{\otimes N}$ & $\times$ & $(NM)^N$ & $|12\rangle|\phi_1\phi_2\rangle$ \\
         \hline
         \makecell{First \\ quantization} & $\mathcal{B}_{1Q}$ & $\checkmark$ & ${N+NM-1\choose N}$ & \makecell{$\frac{1}{\sqrt{2}}(|12\rangle|\phi_1\phi_2\rangle$ \\ $+|21\rangle|\phi_2\phi_1\rangle)$} \\
         \hline
         \makecell{Second  \\ quantization} & $\mathcal{B}_{2Q}$ & $\checkmark$ & ${N+NM-1\choose N}$ & $|10\ 01\rangle\rangle$ \\
         \hline
         \makecell{Mode assignment \\ lists} & $\mathcal{B}_{\text{MAL}}$ & $\times$ & $M^N$ & $|12)$ \\
         \hline
    \end{tabular}
    \caption{A summary of symmetrized and unsymmetrized bases. The example states are all equivalent up to symmetrization.}
    \label{tab:bases}
\end{table}

There are two redundancies that cause this basis to be smaller than the full unsymmetrized basis in \cref{eq:basis-general}. The first is due to the label degree of freedom, which we have combined with the internal mode degree of freedom. The second is that the $N$-particle basis contains states with more than one photon in the same internal mode, such as $|12\rangle|\phi_1\phi_1\rangle$. Using mode assignment lists, it is still possible for multiple photons to have the same internal state, but they have different labels until the end of the computation. By eliminating these two sets of basis states, we get the reduced space size $M^N$.

For example, consider a system with two photons in two modes. The basis elements in second quantization are
\begin{equation}\label{eq:2Q_HOM}
    \begin{split}
        \mathcal{B}_{2Q}=\{&|20\ 00\rangle,|11\ 00\rangle,|10\ 10\rangle,|10\ 01\rangle,|02\ 00\rangle, \\
        &|01\ 10\rangle,|01\ 01\rangle,|00\ 20\rangle,|00\ 11\rangle,|00\ 02\rangle\}.
    \end{split}
\end{equation}
These ten basis states are not affected by exchanging photon labels, but six of them contain two photons in the same internal mode. The mode assignment list basis treats those six states as redundant so it only has four elements,
\begin{equation}\label{eq:MAL_HOM}
    \mathcal{B}_\text{MAL}=\{|11),|12),|21),|22)\}.
\end{equation}
The middle two basis states differ by more than just a label exchange, since the two photons implicitly have different internal states.

\subsection{Unitary operators in the unsymmetrized space}
Working in the unsymmetrized space of mode assignment lists $\mathcal{H}_\text{MAL}$ allows us to store a smaller density matrix. The elements $\{E_i\}$ of the circuit operation \cref{eq:op-sum} will have the same smaller size. This is also true of the unitary operators representing circuit components. Since the photons are effectively distinguishable by their labels, every photon traverses the circuit independently of the other photons regardless of their internal states. This means we can apply circuit components by operating on mode assignment lists with products of single-photon unitaries~\cite{Beggi2018}.

The single-photon unitary $U_1$ representing a particular circuit component exists in the external mode space $\mathcal{H}_\text{ext}$, which has dimensions $M\times M$. The unitary that applies this operation to every photon in the system is
\begin{equation}
    U=U_1^{\otimes N}.
    \label{eq:U}
\end{equation}
For example, consider a beam splitter that operates on two modes, $U(m_1,m_2)$. If the two targeted modes are $1$ and $2$, then the single-photon unitary is~\cite{Campos1989,Demirel2019}
\begin{equation}
    U_1(1,2)=\exp{\left(i\theta\sigma_y\right)}\oplus \mathbb{I}_{M-2},
\end{equation}
where $\sigma_y$ is the Pauli-y matrix and $\theta$ is an angle that determines the beam splitter ratio ($\theta=\pi/4$ is a balanced splitter). The identity is added to leave the photons in the remaining $M-2$ external modes unaffected. This can also be transformed into a beam splitter acting on any two modes $m_i$ and $m_j$ by permuting the rows and columns of $U_1(1,2)$ such that $(1,2)\rightarrow(m_i,m_j)$. This operator acts individually on each of the $N$ photons, giving a unitary of the form \cref{eq:U} with dimensions $M^N\times M^N$.

\section{Resolving interference}
\label{sec:resolve}
\subsection{Tracing out internal modes}
The result of a simulation in the space of mode assignment lists is a density matrix of the form
\begin{equation}
    \mu=\sum_{ij}\mu_{ij}|\bm{m}_i)(\bm{m}_j|.
    \label{eq:rho-mal}
\end{equation}
The density matrix contains the location of each individual photon at the output of the circuit without any interference effects. In a sense, we have put off deciding whether the particles in the system are bosons or fermions until after the circuit operation. In order to implement the bosonic properties of the system so that the photons interfere properly, we apply the symmetrization operator $\mathcal{S}$, defined in \cref{eq:S}. Doing so gives a density matrix in first quantization,
\begin{equation}
    \begin{split}
        \rho_{1Q}&=\mathcal{S}\mu\mathcal{S}^\dagger \\
        &=\frac{1}{|S_N|}\sum_{ij}\mu_{ij}\sum_{\pi,\sigma\in S_N}\frac{\pi|\bm{m}_i)(\bm{m}_j|\sigma^\dagger}{\sqrt{|I_{\bm{m}_i,\bm{\phi}^{(0)}}||I_{\bm{m}_j,\bm{\phi}^{(0)}}|}},
    \end{split}
\label{eq:rho-sym}
\end{equation}
where $i,j$ index the set of possible external states since $|\bm{\phi}^{(0)}\rangle$ is always the same.

The problem with symmetrizing the state is that the resulting state inhabits the space $\mathcal{H}_S$, which can be much larger than $\mathcal{H}_{\text{MAL}}$. To avoid having to use this larger space, we can immediately perform a partial trace over the internal modes without saving the state \cref{eq:rho-sym}. This assumes any measurements we will eventually make on the system will be independent of the internal states of the photons. We will refer to the combined operation of symmetrizing and tracing out internal modes as \emph{resolving the interference}. The result will be an external density matrix~\cite{BrunnerMasters,Stanisic2018,Dittel2021}, which contains no information about the internal states of the photons leaving the circuit. In practice, we are often most concerned with the external mode information such as probabilities of different detection patterns. The resulting external density matrix is
\begin{equation}
    \begin{aligned}
        \rho_{\text{ext}}&=\text{Tr}_\text{int}(\mathcal{S}\mu\mathcal{S}^\dagger) \\
        &=\frac{1}{|S_N|}\sum_{ij}\sum_{\pi,\sigma\in S_N}&&\frac{\mu_{ij}\langle\bm{\phi}^{(0)}|\sigma^\dagger\pi|\bm{\phi}^{(0)}\rangle}{\sqrt{|I_{\bm{m}_i,\bm{\phi}^{(0)}}||I_{\bm{m}_j,\bm{\phi}^{(0)}}|}} \\
        & &&\times\pi|\bm{m}_i\rangle\langle\bm{m}_j|\sigma^\dagger.
    \end{aligned}
    \label{eq:rho-ext-1q}
\end{equation}
The external density matrix exists in the external space $\mathcal{H}^{\otimes N}_\text{ext}$, which has the same size as $\mathcal{H}_{\text{MAL}}$. So by skipping over \cref{eq:rho-sym}, we never have to construct a density matrix in the larger space $\mathcal{H}_S$.

The state after the partial trace is not necessarily symmetric. Although the overall state of a bosonic system must be symmetric, the state of a subsystem, such as the reduced external state, does not have to be symmetric. Since the reduced state is not symmetric, there is no clear way to write it in second quantization without breaking the symmetry~\cite{BrunnerMasters}. The only exception is when the photons are all mutually indistinguishable, in which case the state is symmetric and can be rewritten in an external Fock space.

\begin{figure}
    \centering
    \includegraphics[trim=0 0 0 0, clip, width=0.5\columnwidth]{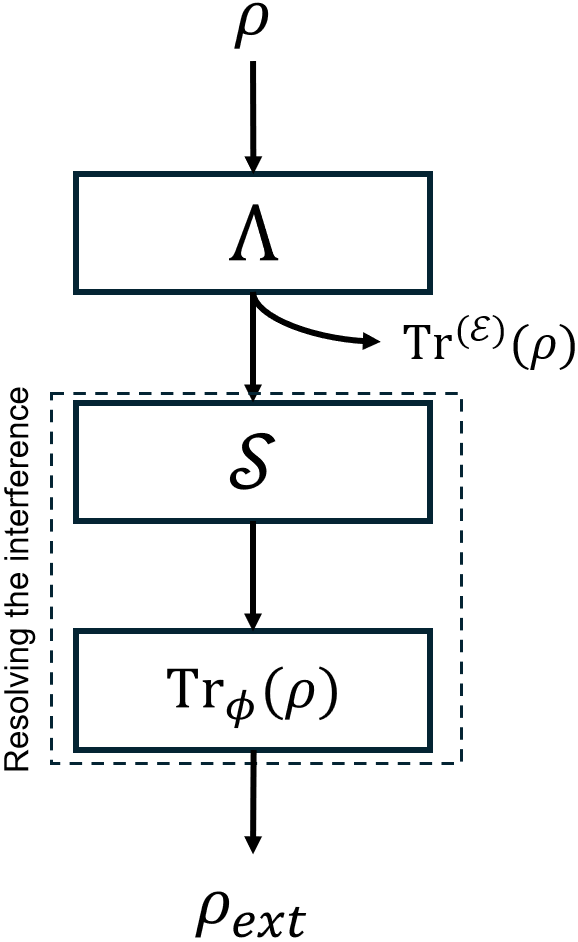}
    \caption{An illustration of the proposed simulation procedure. The state is initially represented in the mode assignment list basis, which means it is not symmetrized. The symmetrization operator $\mathcal{S}$ commutes with operations that act on the external mode space, so we apply the circuit operation $\Lambda$ before symmetrizing. Since the photons are effectively distinguishable in the unsymmetrized space, we use a distinguishable-particle version of the circuit operation. After the circuit, we combine symmetrization with a partial trace over the internal modes, resulting in an external density matrix and skipping over the large symmetrized space $\mathcal{H}_S$. It is also possible, as discussed in \cref{sec:trace-photons}, to only resolve the interference for a subset of photons that were detected by tracing out those photons, $\text{Tr}^{(\mathcal{E})}(\rho)$. The state of the remaining photons can then continue on to another circuit.}
    \label{fig:workflow-2}
\end{figure}

To summarize, the procedure we are proposing for density matrix simulation is to
\begin{enumerate}
    \item express the input state as a density matrix in the basis of mode assignment lists $\mathcal{B}_{\text{MAL}}$,
    \item apply a distinguishable-photon circuit operation as in \cref{eq:U}, and
    \item resolve the interference by symmetrizing and tracing out the internal states as shown in \cref{eq:rho-ext-1q}.
\end{enumerate}
\cref{fig:workflow-2} contains an illustration of the proposed procedure.

\subsection{Detection pattern probabilities}
We define a detection pattern $\bm{d}$ to be a set of occupation numbers in the external space. The detection pattern for an external state $|\bm{m}_{\bm{d}}\rangle$ is independent of particle labelling, so all permutations of $|\bm{m}_{\bm{d}}\rangle$ have the same detection pattern $\bm{d}$. The probability to measure a certain detection pattern $\bm{d}$ is the sum of the probabilities of measuring any of its corresponding permutations,
\begin{equation}
    \begin{split}
        P(\bm{d})&=\frac{1}{|I_{\bm{m}_{\bm{d}}}|}\sum_{\pi\in S_N}\langle\bm{m}_{\bm{d}}|\pi^\dagger\rho_{\text{ext}}\pi|\bm{m}_{\bm{d}}\rangle \\
        &=\sum_{ij}\mu^{(\bm{d})}_{ij}\langle\Phi_j|\Phi_i\rangle,
    \end{split}
    \label{eq:det-prob}
\end{equation}
where $|\bm{m}_{\bm{d}}\rangle$ can be any permutation and $\mu^{(\bm{d})}$ is the submatrix of \cref{eq:rho-mal} containing basis states with the detection pattern $\bm{d}$. We also define
\begin{equation}
    |\bm{\Phi}_i\rangle=\bigotimes_{k=1}^{M}\mathcal{S}|\bm{\phi}^{(0)}_{\mathcal{P}_{k,j}}\rangle.
    \label{eq:phi-star}
\end{equation}
to be the photons' internal states symmetrized within each external mode, where $\mathcal{P}_{k,j}$ is the set of photons in spatial mode $k$ for the state $|\bm{m}_j\rangle$. For example, if $|\bm{m}_i\rangle=|211\rangle$, then $|\Phi_i\rangle=\frac{1}{\sqrt{2}}(|\phi_2\phi_3\rangle+|\phi_3\phi_2\rangle)\otimes|\phi_1\rangle$. \cref{app:det} contains details on how \cref{eq:det-prob} is derived\footnote{In the literature, these same detection pattern probabilities are occasionally derived by tracing out the internal modes using the logical-internal basis~\cite{SparrowThesis,Saied2024}. We comment on this basis and its place in our framework in \cref{app:sparrow}}. It is not necessary to perform the intermediate steps described in this section, where $\mu$ is symmetrized and the internal modes are traced out, in order to obtain the detection probabilities. If the probabilities are the desired result of a simulation, then \cref{eq:det-prob} can be used directly from the output \cref{eq:rho-mal}. 

\subsection{Tracing out spatial modes}
\label{sec:trace-photons}
Some circuits involve a form of post-selection in which a particular subset of external modes are measured, and only certain measurement outcomes are considered a success~\cite{Chabaud2021,Bartolucci2021,Pankovich2023b,Raussendorf2001,Walther2005,Monbroussou2024}. Photons that reach a detector must have their interference resolved in order to give the correct detection probabilities. While it is possible to postpone resolving the interference of all photons until the end of the circuit, it is often beneficial to resolve the interference of certain photons as they are measured. This allows the external modes ending in detectors to be discarded from the simulation, leaving any photons in the undetected modes to continue on through the circuit. If the simulation is limited by memory constraints, then this allows for more photons to be introduced into the circuit after each set of detectors, making the method more scalable.

In order to trace out a subset of external modes $\mathcal{E}$, we symmetrize then trace out the states of the photons in those spatial modes. We define a subset $\mathcal{T}$ of photons to be targeted, which contains every photon with an external state that matches one of the spatial modes being traced; in other words, photon $\ell_i$ is targeted if $m_i\in\mathcal{E}$. We define a partial symmetrization operator that acts on the targeted photons and ignores the others,
\begin{equation}
    \mathcal{S}_{\mathcal{T}}|\bm{m},\bm{\phi}\rangle=\frac{1}{\sqrt{|S_{\mathcal{T}}||I_{\mathcal{T}}|}}\sum_{\pi\in S_{\mathcal{T}}}\pi|\bm{m},\bm{\phi}\rangle,
\end{equation}
where $S_{\mathcal{T}}$ is the symmetric group for the set of targeted photons and $I_{\mathcal{T}}$ is the group of permutations under which the targeted photons' states are invariant.

The density matrix after symmetrizing and tracing out the targeted photons is
\begin{equation}
    \text{Tr}^{(\mathcal{E})}(\mu)=\sum_{\bm{d}}\sum_{ij}\mu_{ij}^{(\bm{d})}|\bm{m}_{i,\notin\mathcal{E}})(\bm{m}_{j,\notin\mathcal{E}}|\langle\bm{\Phi}_{j,\in\mathcal{E}}|\bm{\Phi}_{i,\in\mathcal{E}}\rangle
    \label{eq:rho-detected-2}
\end{equation}
where the notation $\text{Tr}^{(\ldots)}$ is used to indicate a trace over a certain subset of external modes. The states of the undetected photons are still represented as mode assignment lists, so the state can continue on through the simulation in the unsymmetrized space. This result is a mixed state over different detection patterns; terms can be kept or discarded depending on the relevant post-selection criteria. Post-selection is probabilistic, so the post-selected state must be re-normalized to account for the discarded terms.

The same detection pattern can be created by different combinations of photons being detected, so we need to keep track of which photon labels have been erased from each mode assignment list. A practical way to do this is to introduce a fictitious external mode, which we call the $0$th mode, to store photons that are no longer in the circuit. In other words, when a photon leaves the circuit, its entry in the mode assignment list is replaced with $0$. This keeps the labelling order in the mode assignment lists untouched, so the $i$th position in the list still corresponds to the photon with label $\ell_i$ and the corresponding internal state. The presence of this fictitious $0$th external mode increases the upper bound on the mode assignment list space size from to
\begin{equation}
\text{dim}(\mathcal{H}_\text{ext}^{\otimes N})=(M+1)^N,   
\end{equation}
which is still significantly smaller than the size of the symmetrized space $\mathcal{H}_S$.

\section{Loss of partially distinguishable photons}
\label{sec:loss}

\subsection{Loss modes}
Loss in a linear optical circuit can be modelled as if lost photons were being moved into a fictitious, inaccessible external mode called a \emph{loss mode}~\cite{NielsenBook,Dorner2009}. This is often how it is done in existing circuit simulations~\cite{SOQCS,Saied2024} because it keeps the state pure. Tracing out the loss modes results in a mixed state. A new loss mode must be introduced each time loss is applied to the circuit so that photons that are already lost do not re-enter the system. There must also be a different loss mode for each system mode, which ensures that when the loss modes are traced out, each type of loss becomes a different term in the mixture. The presence of loss modes means that simulating a system with loss requires a much larger space. This problem is exacerbated when the photons are not identical since the number of system modes becomes $N\times M$ in both first and second quantization, where each system mode is a specific combination of external and internal modes. Each of these combinations must have a different loss mode. If the chosen simulation method is restricted to pure states, which is the case for matrix permanent simulations, all the loss modes must be retained until the end of the circuit because tracing them out results in a mixed state.

In existing work, this loss model is often simplified by assuming each loss element implements uniform loss across all modes. The uniform loss model allows each loss element to be commuted through other components in the circuit~\cite{Oszmaniec2018, Saied2024}. In first quantization, each possible outcome of a loss event is a trace over a subset of $n\leq N$ photons, which commutes with unitary operations
\begin{equation}
    \text{Tr}^{(n)}(U_1^{\otimes N}\rho_{1Q} (U_1^\dagger)^{\otimes N})=U_1^{\otimes (N-n)}\text{Tr}^{(n)}(\rho_{1Q})(U_1^\dagger)^{\otimes (N-n)}.
    \label{eq:commute-loss}
\end{equation}
Regardless of the number of photons remaining in the system, each photon sees $U_1$, as described in \cref{eq:U}. Since $\rho_{1Q}$ is in first quantization, which is symmetric, this result is independent of which photon is traced out.

Commuting uniform loss through circuit components can lead to a simpler simulation by gathering multiple loss elements together and combining their loss rates. If all the loss in the system can be commuted to the beginning or the end of the circuit, effectively restricting loss to the sources and detectors, then there are further results that can be used to simplify simulations of common entanglement-generation circuits~\cite{Varnava2008}.

\subsection{Kraus operators}
In a density matrix simulator, loss can be handled more naturally. Density matrices can store mixed states, so the circuit operation does not need to be unitary. In the workflow in \cref{fig:workflow-2}, loss takes place as part of the circuit operation $\Lambda$. Each loss mode is effectively traced out at the same time as it is introduced, resulting in a map of the form \cref{eq:op-sum} that keeps the system the same size. Loss is applied using a set of Kraus operators $\{K_i\}$, which are applied to the density matrix according to
\begin{equation}
    \Lambda_{\text{loss}}(\rho)=\sum_i K_i\rho K_i^\dagger.
\end{equation}
This creates a mixed state where each term corresponds to a different possible outcome of a loss event.

For indistinguishable photons in a single external mode, the possible outcomes are to lose any number $n\leq N$ of the photons. The Kraus operators for indistinguishable photons are~\cite{Dorner2009,Demkowicz2014}
\begin{equation}
    K_n=\eta^{\frac{N-n}{2}}(1-\eta)^{\frac{n}{2}}\frac{a^n}{\sqrt{n!}},
    \label{eq:kraus-indist}
\end{equation}
where $a$ is the photon annihilation operator and $\eta$ is the probability of a photon being lost. For perfectly distinguishable photons, each type of photon undergoes loss independently from photons of other types. The Kraus operators acting on the system are products of Kraus operators for each photon type, each using a different annihilation operator $a_i$. The total Kraus operators are
\begin{equation}
    K_{\bm{n}}=\eta^{\frac{N-n}{2}}(1-\eta)^{\frac{n}{2}}\prod_{i}\frac{a_i^{n_i}}{\sqrt{n_i!}},
    \label{eq:kraus-dist}
\end{equation}
where $\bm{n}$ is a list containing the number of lost photons of each type.

Loss is more complicated when the photons are partially distinguishable due to the non-orthogonal internal states. Each annihilation operator $a(\phi_i)$ acts on photons with the corresponding internal state $\phi_i$, but will also affect photons with different internal states if there is a non-zero overlap. These operators obey the commutation relation~\cite{Shchesnovich2022}
\begin{equation}
    [a(\phi_i),a^\dagger(\phi_j)]=S_{ij}\mathbb{I},
\end{equation}
where $S_{ij}=\langle\phi_i|\phi_j\rangle$ is the overlap between the two internal states. It is possible to perform a linear transformation to orthogonalize this set of annihilation operators,
\begin{equation}
    \{a(\phi_i)\}\rightarrow\{a(\tilde{\phi}_i)\},
\end{equation}
where $\langle\tilde{\phi}_i|\tilde{\phi}_j\rangle=\delta_{ij}$ (see \cref{app:gs}). Each of the orthogonalized operators acts on photons with a specific orthogonalized internal state without affecting any other photon types. Working in this basis allows us to use an orthogonalized version of the distinguishable-photon Kraus operators,
\begin{equation}
    K_{\bm{n}}=\eta^\frac{N-n}{2}(1-\eta)^\frac{n}{2}\prod_{i}\frac{a^{n_i}(\tilde{\phi}_i)}{\sqrt{n_i!}},
    \label{eq:kraus}
\end{equation}
where $\bm{n}$ now contains the number of lost photons with each orthogonalized internal state. The resulting state has no dependence on the choice of orthogonal basis. A more detailed derivation of these Kraus operators can be found in \cref{app:loss}. These Kraus operators reduce to either \cref{eq:kraus-indist} or \cref{eq:kraus-dist} in the corresponding edge cases of identical or perfectly distinguishable photons.

For example, consider a system of two partially distinguishable photons,
\begin{equation}
    |\psi\rangle=|11\rangle\rangle=\frac{a^\dagger(\phi_1)a^\dagger(\phi_2)}{\sqrt{1+|S_{12}|^2}}|\text{vac}\rangle\rangle,
\end{equation}
where the factor in the denominator keeps the state normalized. The Kraus operators are $K_{(0,0)},K_{(1,0)},K_{(0,1)},K_{(2,0)},K_{(1,1)},K_{(0,2)}$, where the subscripts indicate the number of lost photons with internal states $\tilde{\phi}_1$ and $\tilde{\phi}_2$. Applying these Kraus operators to the state gives
\begin{equation}
    \begin{split}
        &\eta^2 |11\rangle\rangle\langle\langle 11|+(1-\eta)^2|00\rangle\rangle\langle\langle00| \\
        +&\frac{\eta(1-\eta)}{1+|S_{12}|^2}\left(|01\rangle\rangle\langle\langle01|+|10\rangle\rangle\langle\langle10|\right) \\
        +&\frac{S_{12}\eta(1-\eta)}{1+|S_{12}|^2}\left(|01\rangle\rangle\langle\langle10|+|10\rangle\rangle\langle\langle01|\right),
    \end{split}
\end{equation}
which reduces to the expected results when $S_{12}=0$ or $1$. Note that this operation is trace-preserving, which can be seen by taking the trace using a basis of orthogonal internal states.

It is still possible to commute this type of loss through a unitary component as long as the loss is uniform across all targeted modes. Just like when the photons are identical, each outcome of the loss corresponds to a trace over a subset of photons. The only difference is that the trace is now in the orthogonalized basis. Since the trace only affects lost orthogonalized photons, we can use the same reasoning as \cref{eq:commute-loss} to show that uniform loss commutes with unitary circuit components.

\subsection{Loss using mode assignment lists}
Consider applying loss to a targeted external mode with loss probability $1-\eta$. There is a set $\mathcal{T}$ of photons in the targeted mode and it is possible for any subset $\mathcal{L}_n$ of $n\leq|\mathcal{T}|$ photons to be lost. The map that applies this operation to a density matrix in the mode assignment list basis is
\begin{equation}
    \begin{split}
        \Lambda_{\text{loss}}(\mu)&=\sum_{ij}\mu_{ij}\sum_{n=0}^{\text{min}(|\mathcal{T}_i|,|\mathcal{T}_j|)}\eta^{\frac{|\mathcal{T}_i|+|\mathcal{T}_j|}{2}-n}(1-\eta)^n \\
        &\times \sum_{\substack{\mathcal{L}_{n_i}\in\mathcal{T}_i,\\ \mathcal{L}_{n_j}\in\mathcal{T}_j}}\langle\bm{\Phi}_{\mathcal{L}_{n_i}}|\bm{\Phi}_{\mathcal{L}_{n_j}}\rangle|\bm{m}_{i,\notin\mathcal{L}_{n_i}})(\bm{m}_{j,\notin\mathcal{L}_{n_j}}|.
    \end{split}
    \label{eq:lossmap}
\end{equation}
where $|\bm{m}_{i,\notin\mathcal{L}_n})$ is the mode assignment list containing the photons that were not lost. The internal states $|\bm{\Phi}_{\mathcal{L}_n}\rangle$ have the form \cref{eq:phi-star}, but only include the internal states of the lost photons in $\mathcal{L}_n$. A full derivation of this result can be found in \cref{app:loss}.

This map considers every possible combination of lost photons then resolves the interference for only the lost photons, symmetrizing before tracing them out. In practice, lost photons can all be assigned a label of $0$, just like detected photons, to keep track of which photon was lost. The map in \cref{eq:lossmap} mixes the state at the point where the loss occurs, so the only reason to have an extra mode at all is to keep track of which photon was lost.

\section{Results}
\label{sec:examples}
\subsection{Hong-Ou-Mandel experiment}
\label{sec:examples-hom}
In a Hong-Ou-Mandel (HOM) experiment~\cite{HOM1987,Branczyk2017}, one photon enters each port of a balanced beam splitter. When the photons are perfectly indistinguishable, every outcome of the experiment should see both photons arrive in the same external mode. This effect is called bunching, and generalizations to higher numbers of modes and partially distinguishable photons have recently drawn some interest~\cite{Pioge2024,Seron2023,Robbio2024}. As the photons become more distinguishable, the probability of detecting a coincidence event, where one photon arrives in each of the external modes, grows. In this section, we analytically demonstrate our method on a standard two-photon HOM experiment.

The input state to a HOM experiment is the mode assignment list
\begin{equation}
    |\bm{m})=|12).
\end{equation}
We choose $|\bm{\phi}^{(0)}\rangle=|\phi_1\phi_2\rangle$, which means that the photon in the first position has internal state $\phi_1$ and the second photon has internal state $\phi_2$. Instead of symmetrizing the state, we will immediately apply a single-photon beam splitter operation to each photon as in \cref{eq:U}. Using the unitary
\begin{equation}
    U=\left(e^{i\frac{\pi}{4}\sigma_y}\right)^{\otimes 2}=\frac{1}{2}\begin{pmatrix} 1 & 1 \\ -1 & 1 \end{pmatrix}\otimes \begin{pmatrix} 1 & 1 \\ -1 & 1 \end{pmatrix}
\end{equation}
gives
\begin{equation}
    U|\bm{m})=\frac{-|11)+|12)-|21)+|22)}{2}.
\end{equation}
We will focus on calculating the coincidence probability, which is the signature of the HOM experiment. The relevant part of the output density matrix is
\begin{equation}
    \frac{|12)(12|-|12)(21|-|21)(12|+|21)(21|}{4}.
\end{equation}
Note that this result has tracked where each of the input photons ended up after the circuit operation, as if the photons had gone through the circuit without interacting. To implement the overlapping internal states, we resolve the interference by symmetrizing the state and tracing out the internal modes. Resolving the interference using \cref{eq:rho-ext-1q} gives the external density matrix
\begin{equation}
    \frac{1-|\langle\phi_1|\phi_2\rangle|^2}{4}\bigg[|12\rangle\langle12|+|21\rangle\langle21|-|12\rangle\langle21|-|21\rangle\langle12|\bigg].
\end{equation}
The coincidence probability is the trace over all permutations corresponding to one photon in each external mode. Using \cref{eq:det-prob}, we get the familiar HOM result
\begin{equation}
    P(11)=\frac{1-|\langle\phi_1|\phi_2\rangle|^2}{2}.
\end{equation}
When the photons are indistinguishable, the coincidence probability is zero, so the photons always end up in the same external mode. By performing this calculation in the mode assignment list basis instead of the Fock basis, we managed to reduce the basis size from $\text{dim}(\mathcal{H}_S)=10$ to $\text{dim}(\mathcal{H}_{\text{MAL}})=4$, as shown in \cref{eq:2Q_HOM} and \cref{eq:MAL_HOM}.

\subsection{Bell state generator}
The benefit of working in an unsymmetrized basis becomes clearer when dealing with larger circuits. A Bell state generator (BSG) is a fundamental source of entanglement in certain linear optical quantum computing architectures~\cite{Pankovich2023a, Pankovich2023b, Bartolucci2021, Bartolucci2023}. Bell states are a set of entangled states of two qubits of the form
\begin{equation}
    \begin{split}
        |\Phi^\pm\rangle&=\frac{|1010\rangle\rangle\pm|0101\rangle\rangle}{\sqrt{2}} \\
        |\Psi^\pm\rangle&=\frac{|1001\rangle\rangle\pm|0110\rangle\rangle}{\sqrt{2}},
    \end{split}
\end{equation}
where we are using dual-rail qubits as the logical basis, $\{|01\rangle\rangle,|10\rangle\rangle\}$.

\begin{figure}
    \centering
    \includegraphics[trim = 0 0 0 0, clip, width=0.9\columnwidth]{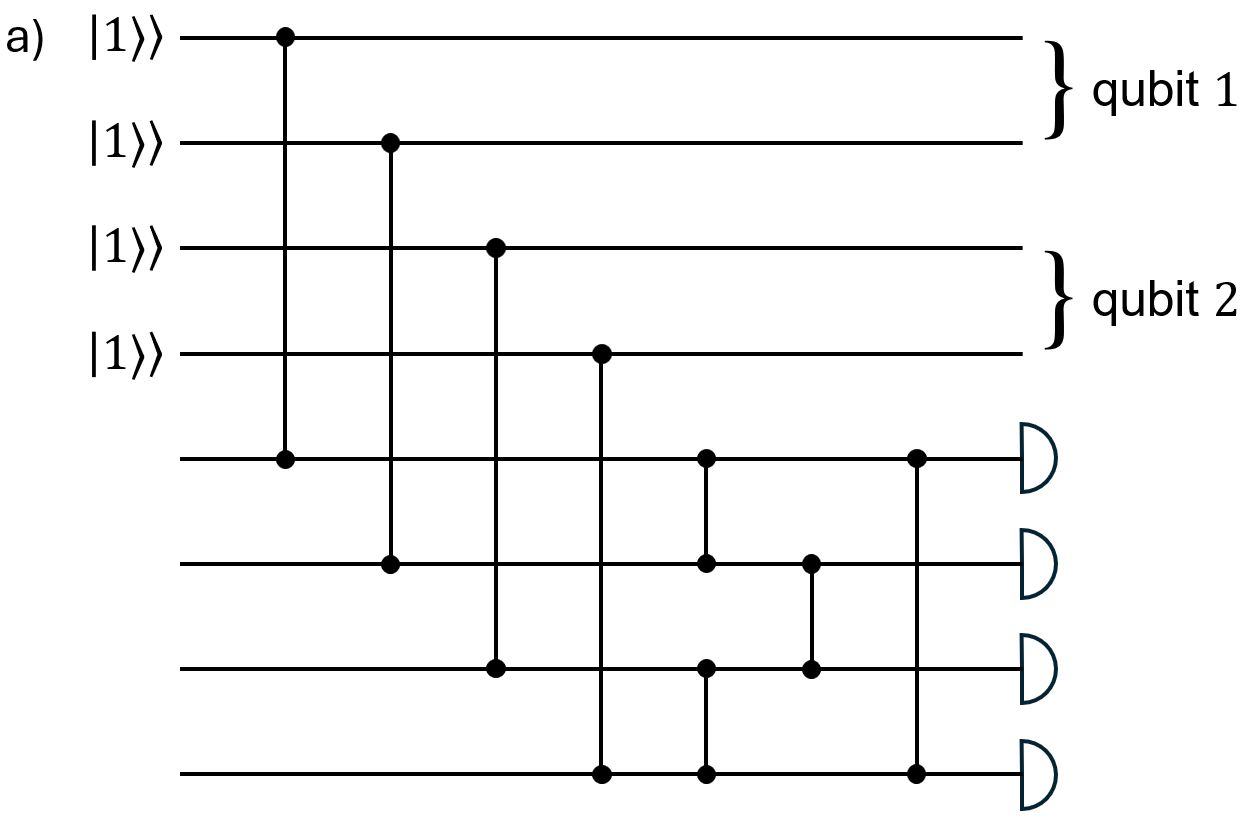} \\
    \vspace{10pt}
    \includegraphics[trim = 0 0 0 0, clip, width=0.9\columnwidth]{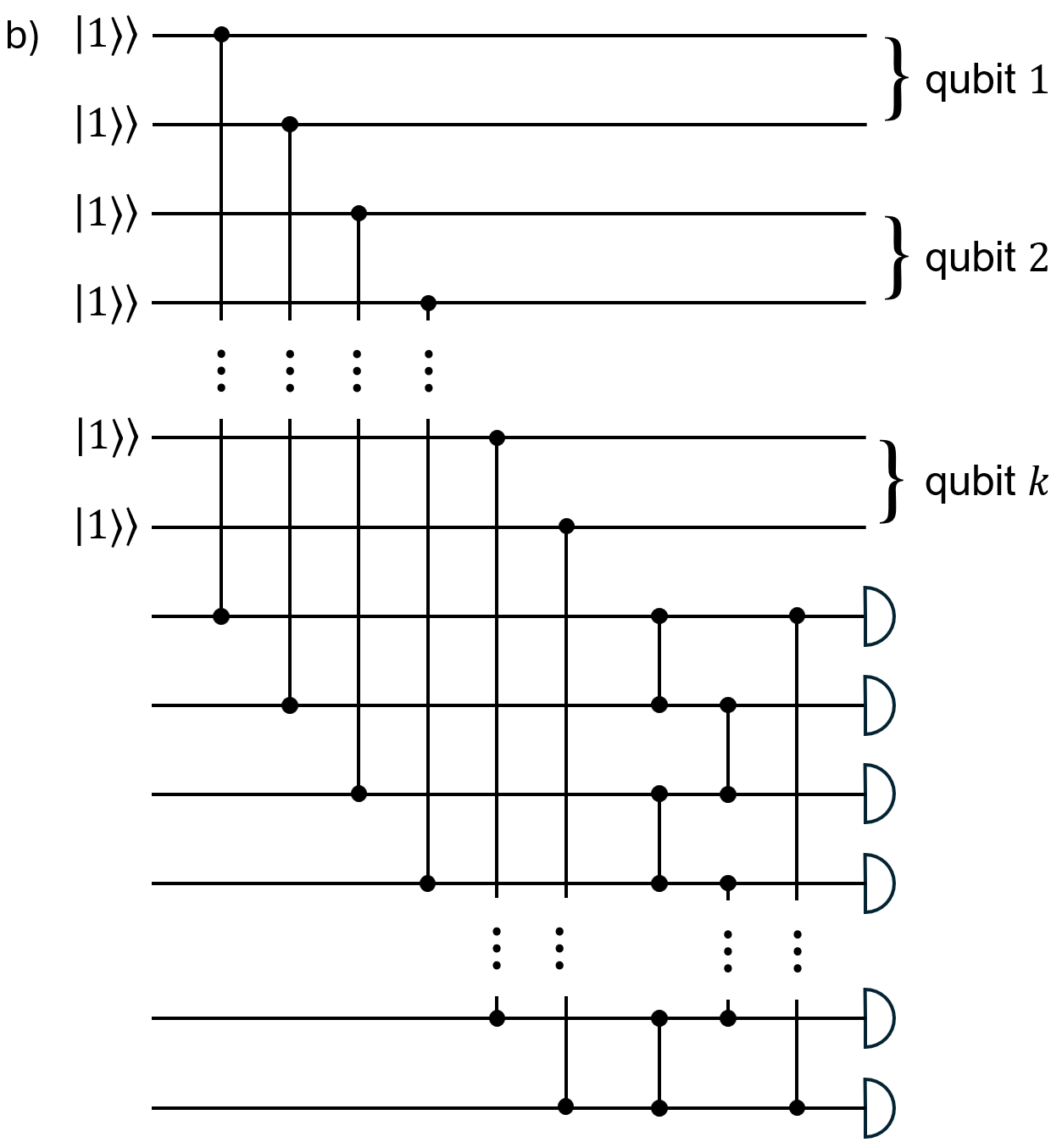}
    \caption{a) A Bell state generator circuit. A Bell state is produced if the detection pattern is any of $(1,1,0,0)$, $(0,0,1,1)$, $(1,0,1,0)$, and $(0,1,0,1)$. Two additional detection patterns, $(1,0,0,1)$ and $(0,1,1,0)$, produce an entangled state that can be converted to a Bell state by swapping modes $2$ and $3$. Each of these patterns occurs with probability $1/32$, so the total probability of a successful herald is $3/16$. b) A generalized $k$-GHZ state generator, which produces an entangled state of $k$ qubits using $2k$ photons in $4k$ external modes.}
    \label{fig:bsg}
\end{figure}

Bell states can be generated using the circuit in \cref{fig:bsg}a~\cite{Zhang2008,Stanisic2017}, with different detection patterns heralding the successful generation of different Bell states. The circuit has four input photons and eight external modes. If the four photons have linearly independent internal states, then the dimension of the symmetrized space is $\text{dim}(\mathcal{H}_S)=52360$, which is the number of ways to distribute the $4$ photons among the $8\times4$ modes, including every pairing of an external mode and an internal mode. The size of the space can be significantly reduced by working in the unsymmetrized space of the mode assignment lists, which has dimension $\mathcal{H}_\text{MAL}=8^4=4096$.

In practice, this number can be reduced even further by restricting the space based on the circuit properties. In the BSG circuit, each photon has a restricted set of external modes where it could possibly be located. For example, the photon initially in external mode $1$ can never end up in external modes $2$, $3$, or $4$ since the order of the beam splitters does not allow those transitions. Extending the same reasoning to the other photons in the circuit, we see that each photon has $5$ allowed external modes, so the restricted space size is $\mathcal{H}_\text{MAL}=5^4=625$. We refer to this kind of space size reduction as using the connectivity of the circuit, and \cref{app:connectivity} covers it in detail. 

\begin{figure}
    \centering
    \includegraphics[trim=20 0 40 40, clip, width=\columnwidth]{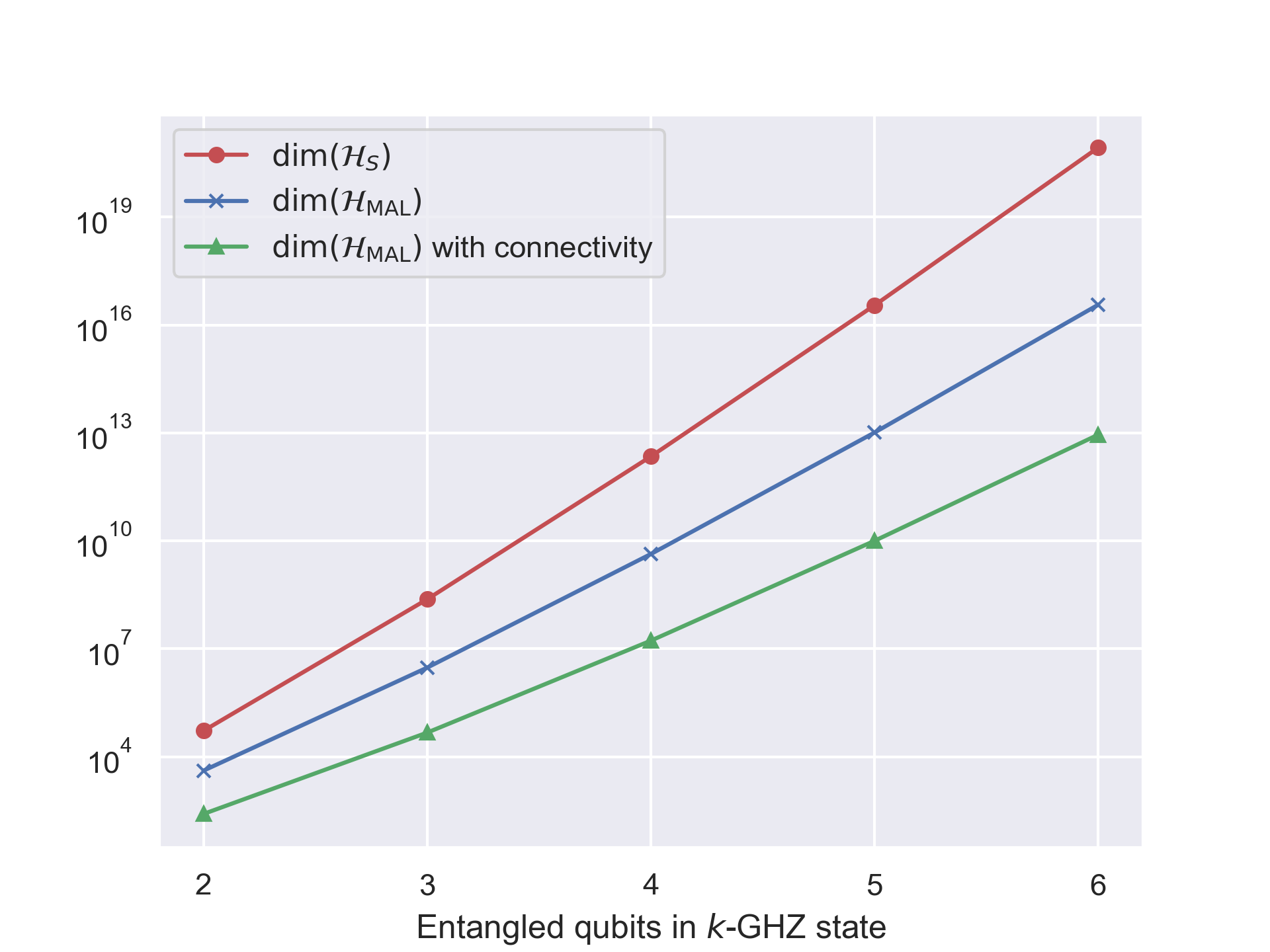}
    \caption{The Hilbert space size for a $k$-GHZ generator, which produces an entangled state of $k$ dual-rail qubits. The case $k=2$ corresponds to a Bell state. The generating circuits are shown in \cref{fig:bsg}. The size scales better using mode assignment lists than it does in Fock space, and can be reduced further by using the connectivity of the circuit. In these Hilbert spaces, every photon in the circuit has a different, arbitrary internal state.}
    \label{fig:bsgscaling}
\end{figure}

The same method can be applied to GHZ states, which are analogous to Bell states but with higher numbers of entangled qubits. \cref{fig:bsg}b shows the generating circuit for arbitrary GHZ states~\cite{Varnava2008,SparrowThesis,GimenoThesis}, which requires $2k$ photons to generate an entangled state of $k$ qubits. The advantage of simulating this circuit using mode assignment lists becomes more significant as more partially distinguishable photons are added to the circuit. \cref{fig:bsgscaling} shows the scaling of the Hilbert space size with the number of entangled qubits in the GHZ state. The scaling is shown for the symmetrized basis $\mathcal{H}_S$, and the (unsymmetrized) mode assignment list basis $\mathcal{H}_\text{MAL}$ with and without using the circuit connectivity.

\begin{figure}
    \centering
    \includegraphics[trim = 0 0 0 30, clip, width=\columnwidth]{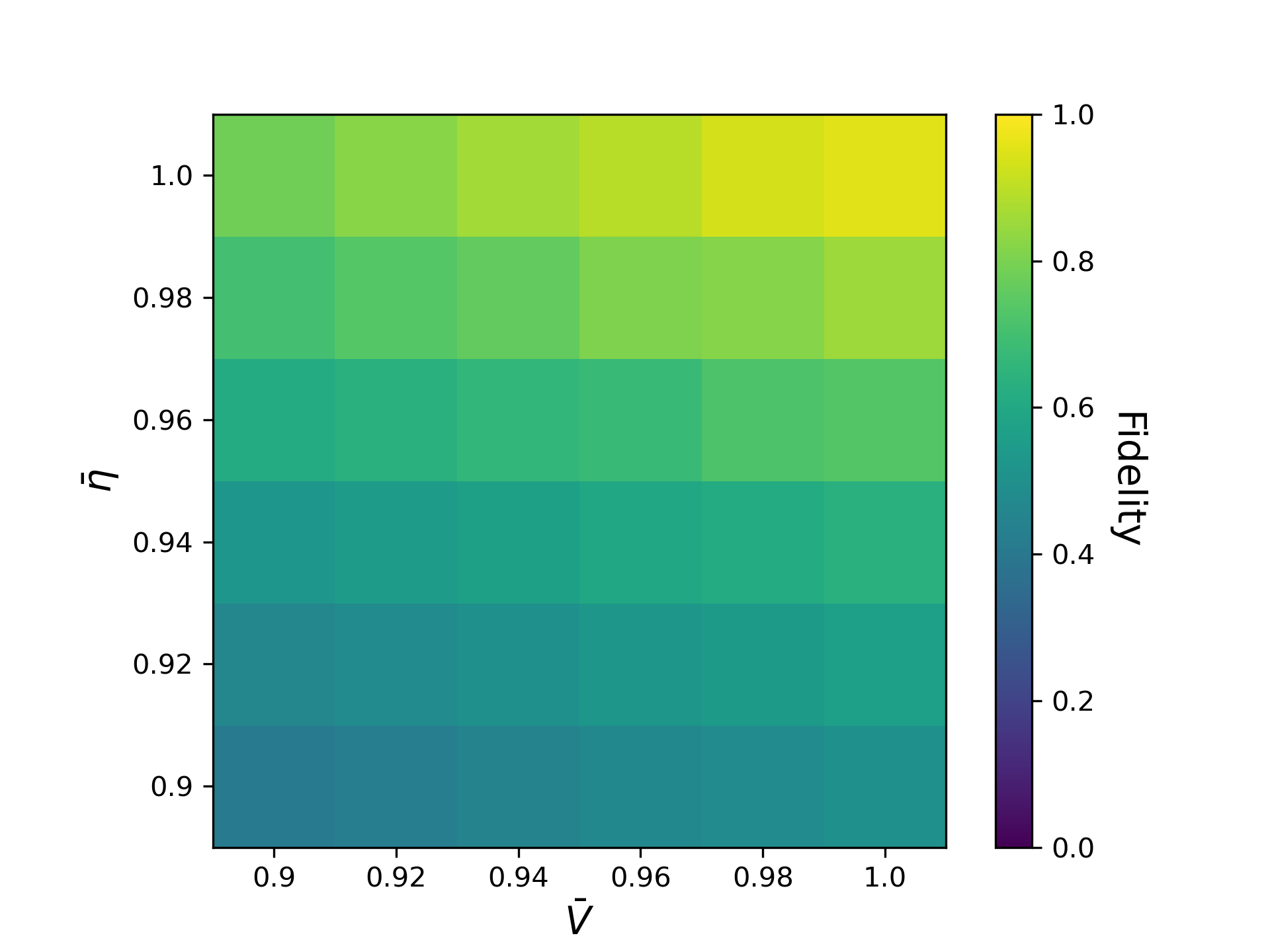}
    \includegraphics[trim = 0 0 0 30, clip, width=\columnwidth]{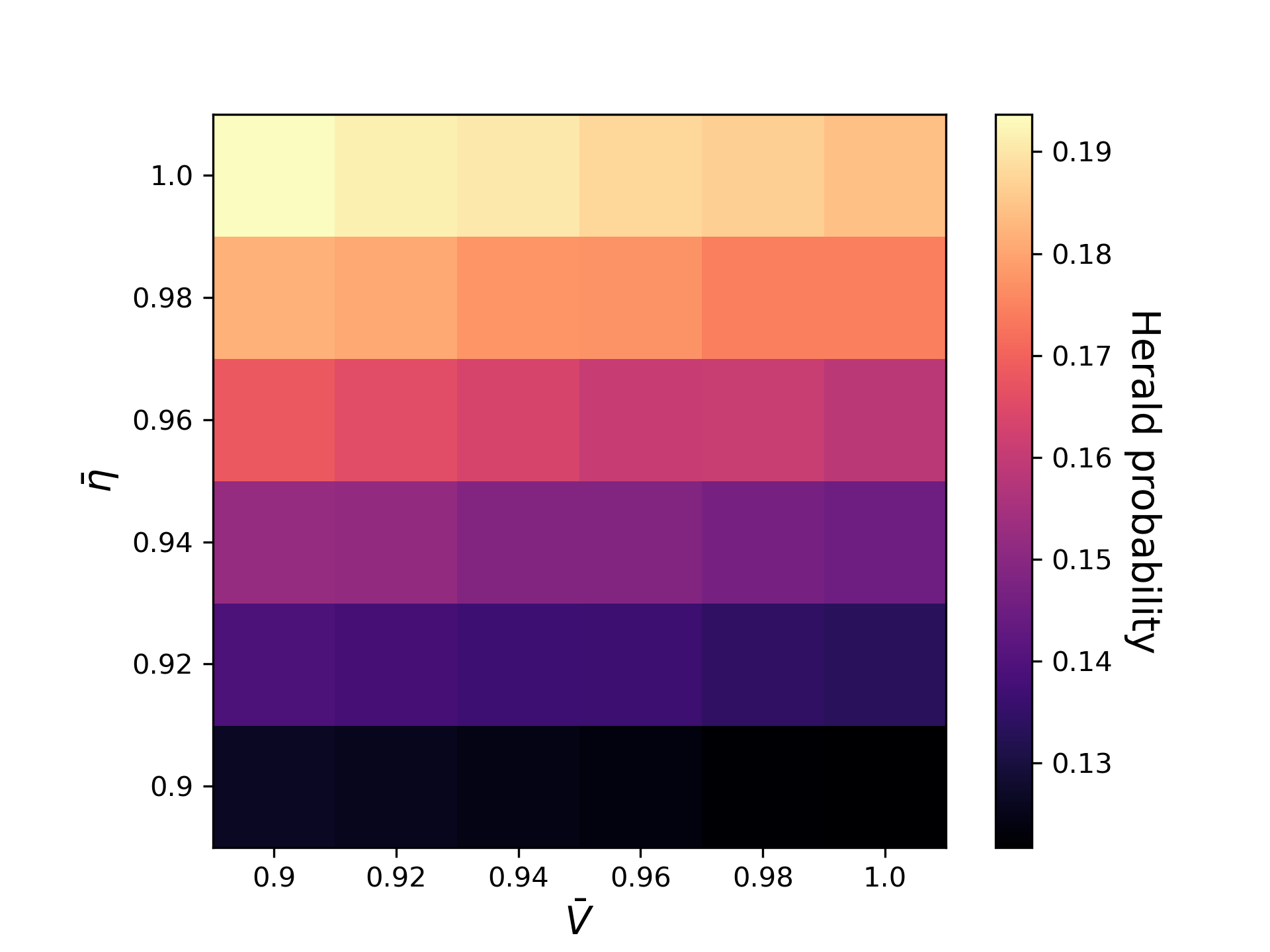}
    \caption{Simulation of a Bell state generator with partially distinguishable photons and loss, showing the external state fidelity and probability of obtaining a successful herald. Each pair of photons has a different visibility and each beam splitter has a different loss coefficient. The pairwise visibilities and loss coefficients are drawn from normal distributions with means $\bar{V}$ and $\bar{\eta}$ and variances $\sigma_V^2=\sigma_\eta^2=10^{-4}$. The resulting fidelity and probability of sucessful Bell state generation are averaged over $10$ simulations, resulting in these distributions. The fidelity \cref{eq:fid} is calculated in the first-quantized basis and uses a Bell state generator with no imperfections as the reference. If there were no errors in the circuit, the herald probability would be $3/16$.}
    \label{fig:bsgresult}
\end{figure}

Working in the mode assignment list basis allows us to simulate both arbitrary partial distinguishability and arbitrary loss without significantly increasing the space size. These two error types can be simulated at the same time using the techniques presented in \cref{sec:loss}. As a demonstration, we have simulated a Bell state generator where every pairwise visibility $|S_{ij}|^2$ and every beam splitter transmission probability $\eta$ is randomly drawn from a normal distribution. This kind of error model would be challenging to simulate using existing methods, which are often restricted to uniform loss and simplified models of distinguishability such as the ``orthogonal bad bit" (OBB) model~\cite{SparrowThesis}, where all pairwise visibilities are the same. \cref{fig:bsgresult} shows the results of this simulation, which outputs an imperfect Bell state density matrix for a range of error parameters. As part of this simulation, we calculate the external state fidelity between the ideal output $\rho_{\text{ext}}$, which is pure, and the imperfect output $\rho'_{\text{ext}}$. Since the ideal state is pure, we use a simplified expression for fidelity,
\begin{equation}
    F(\rho,\rho')=\text{Tr}(\rho\rho').
    \label{eq:fid}
\end{equation}
The fidelity of the output state decreases with both loss and visibility, but is affected more heavily by loss. The probability of measuring a successful herald actually increases from the ideal value of $3/16$ as the photons become more distinguishable, but the state being heralded gets farther away from being a Bell state. If we were to set $\bar{V}=0$, the photons would behave like classical particles, resulting in a product state heralded with $1/4$ probability.

\subsection{QPC-encoded qubit states}
A quantum parity code (QPC) is a way of encoding logical qubits to be more resistant to loss errors and Pauli errors~\cite{Ralph2005, Pankovich2023a, Pankovich2023b}. The QPC$(n,m)$ encoding is a concatenation of two repetition codes on top of the dual-rail encoding,
\begin{equation}
    \begin{split}
        |\bar{0}^{(n,m)}\rangle&=\left[\frac{1}{\sqrt{2}}\left(|10\rangle\rangle^{\otimes m}+|01\rangle\rangle^{\otimes m}\right)\right]^{\otimes n} \\
        |\bar{1}^{(n,m)}\rangle&=\left[\frac{1}{\sqrt{2}}\left(|10\rangle\rangle^{\otimes m}-|01\rangle\rangle^{\otimes m}\right)\right]^{\otimes n}.
    \end{split}
    \label{eq:qpc}
\end{equation}
Larger QPC encodings are more resistant to loss errors, but they are also more difficult to generate, requiring more resources and more complicated circuits. This means they are also more difficult to simulate. The effect of partial distinguishability on the generation and measurement of photonic QPC-encoded states has not previously been quantified.

A QPC$(n,m)$-encoded qubit can be generated in a loss-detecting scheme from $nm$ Bell pairs~\cite{Pankovich2023b}. If each Bell pair is produced using the Bell state generator in Fig.~\ref{fig:bsg}, the QPC-generating circuit uses $4nm$ photons, each of which can have a different internal state, and $8nm$ external modes. Consider the QPC$(4,2)$ encoding, which is generated using a circuit with $N=32$ photons and $M=64$ external modes. If all the photons have different internal states, then a brute force simulation in Fock space has size $\text{dim}(\mathcal{B}_N^{2Q})={2079\choose 32}\sim10^{70}$.

\begin{figure}
    \centering
    \includegraphics[trim = 0 0 0 0, clip, width=\columnwidth]{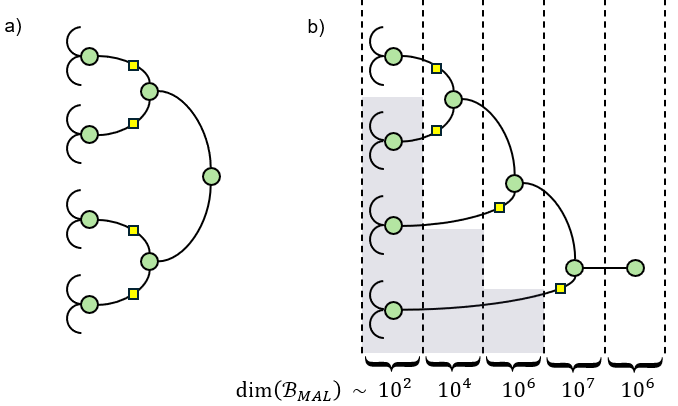}
    \caption{a) A ZX diagram of a QPC$(4,2)$-encoded qubit in the $|+^{(4,2)}\rangle$ state, which can be be converted to a circuit involving Bell state generators, linear optical elements and single photon detectors~\cite{Pankovich2023b}. b) The same circuit redrawn in a cascaded layout, so that new photons are introduced after unconnected detectors are applied. The approximate size of the system is shown for each stage of the circuit. At the end of each stage, a subset of photons and external modes are traced out, freeing up memory for new photons and external modes to be introduced in later stages. Everything in the shaded region can be simulated separately then merged into the cascaded circuit.}
    \label{fig:qpc}
\end{figure}

To reduce the space size, we can delay symmetrization and use the mode assignment list basis to reduce the system size. In the practical implementation, we also use circuit connectivity and a cascaded circuit layout  to further reduce the space size, a process that is described in \cref{app:connectivity}. The original circuit and the cascaded version are shown in Fig.~\ref{fig:qpc}. Both circuits are equivalent, but the cascaded version is broken up into stages where at the end of each stage, unneeded external modes are removed from the system before adding in new ones. Using these techniques, it is possible to perform a density matrix simulation of this circuit using a maximum space size of $\text{dim}(\mathcal{B}_N^{MAL})=1^{10}\times2^{18}\times3^2\times4^2\sim 10^7$. The factors in this product, each of which has the form $M^N$ as in \cref{eq:size-mal}, show the number of external modes (including the fictitious $0$th mode) that each of the $32$ photons can occupy in the largest stage of the cascaded circuit. There are $10$ photons that can only be in one mode, $18$ photons that can be in a maximum of two modes, and so on. For more details on how to obtain this value, see \cref{app:connectivity}.

Reducing the dimension of the Hilbert space from $10^{70}$ to $10^7$ makes it possible to simulate this circuit using a density matrix when normally it would be impractical. Not only can all the photons be partially distinguishable with arbitrary pairwise visibilities, but non-uniform loss can be added throughout the circuit with no additional memory requirements.

\begin{figure}
    \centering
    \includegraphics[trim = 50 0 30 30, clip, width=0.8\columnwidth]{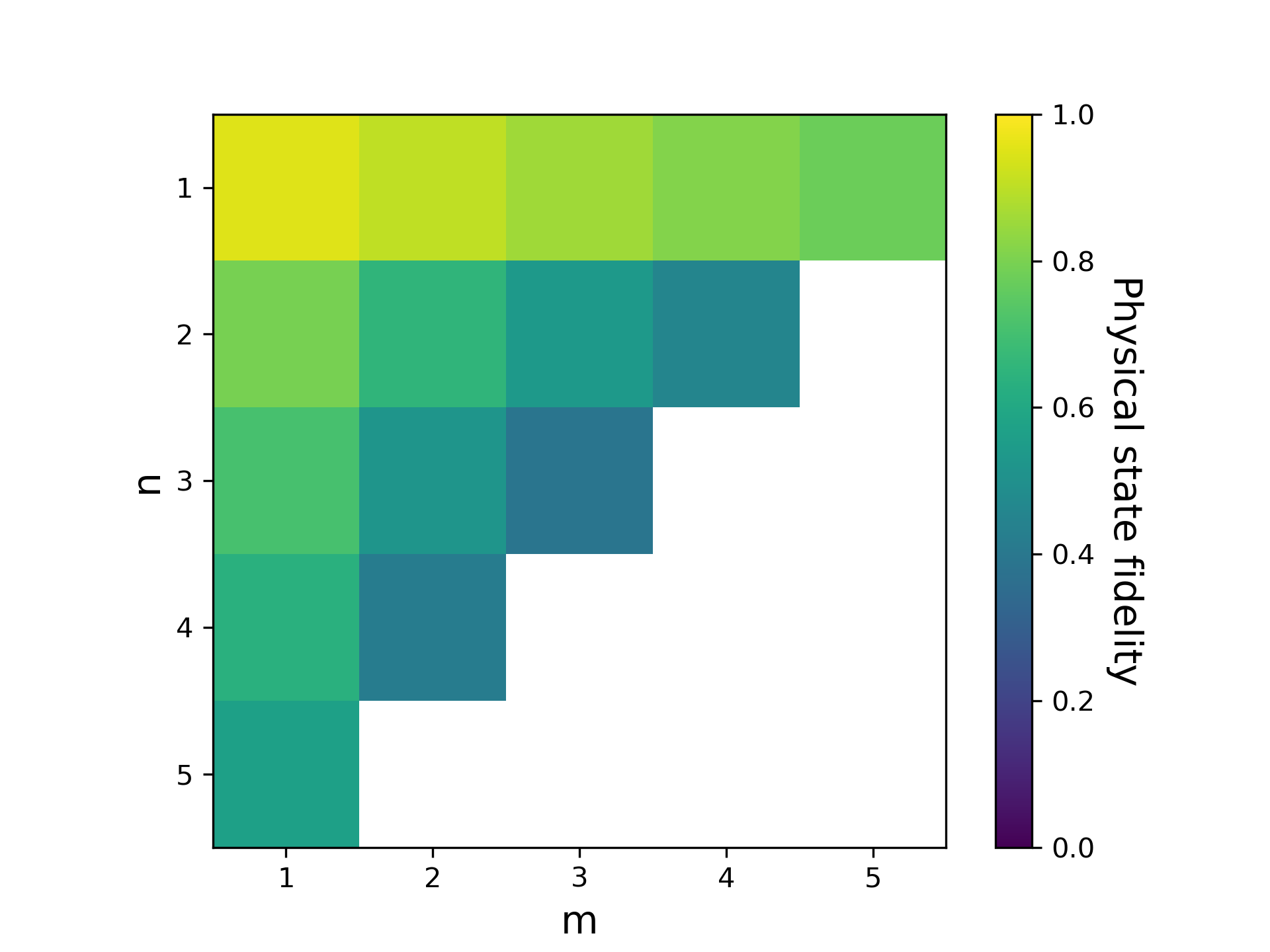}
    \includegraphics[trim = 50 0 30 30, clip, width=0.8\columnwidth]{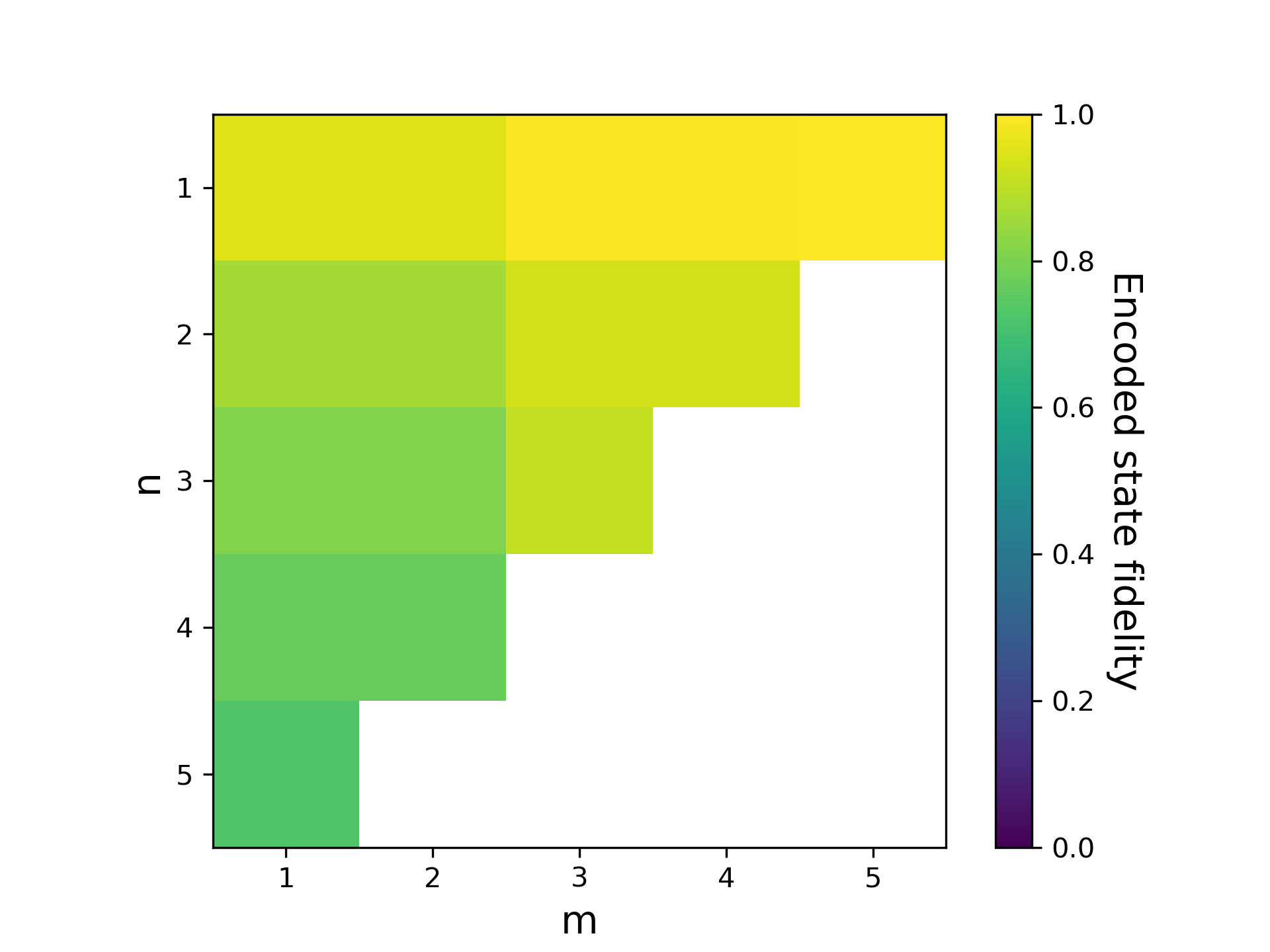}
    \caption{a) Physical state fidelity of a QPC$(n,m)$-encoded qubit in the $|\bar{+}^{(n,m)}\rangle$ state, produced using partially distinguishable photons. Each pair of photons has the same visibility $V=0.9$, although giving each pair a different visibility would not increase the simulation time. For each QPC encoding, we run a simulation to generate a density matrix in the MAL basis and, using the external state fidelity \cref{eq:fid}, compare it against the ideal state produced using indistinguishable photons. Many of these density matrices would be impractical to produce using a Fock space simulation. For example, the QPC(4,2) simulation would normally require a Hilbert space with dimension $10^{70}$, but using mode assignment lists we reduce the space size to $10^7$. b) Encoded state fidelity for the same QPC-encoded state. The encoded state, which in this case is an $X$ eigenstate, gets closer to the ideal state as $m$ increases. When the encoded state is a $Z$ eigenstate instead of $X$, then its encoded fidelity increases with $n$ instead. This is a consequence of the QPC encoding being a concatenation of two repetition codes, as in \cref{eq:qpc}.}
    \label{fig:qpcresult}
\end{figure}

As a demonstration, we have simulated the fidelity of various QPC-encoded single-qubit states using photons that are all partially distinguishable from one another. For the sake of an instructive example, we use the OBB model, where all the pairwise visibilities are the same. Using our method, no additional complexity would be introduced if the pairwise visibilities were all different. The results, which are shown in Fig.~\ref{fig:qpcresult}, show the dependence of QPC$(n,m)$ external state fidelity on the parameters $n$ and $m$. Note that this is the fidelity of the physical state, not the encoded state, so it does not indicate how well the encoding protects the logical information. Larger QPC encodings require larger generating circuits with more instances of photon interference, so the partial distinguishability has more of a negative effect. We include the physical fidelity plot to demonstrate that it is possible to perform calculations that require the physical density matrix.

To see how the encoded state responds to distinguishability errors, we can make encoded measurements on the output state, which involves combining the results of individual physical qubit measurements~\cite{Pankovich2023a}. By measuring the expectation value of each of the encoded observables $\{\bar{X},\bar{Y},\bar{Z}\}$, we can construct a new density matrix,
\begin{equation}
    \bar{\rho}=\frac{1}{2}\left(\mathbb{I}+\langle \bar{X}\rangle\sigma_x+\langle \bar{Y}\rangle\sigma_y+\langle \bar{Z}\rangle\sigma_z\right).
\end{equation}
This density matrix contains all the information that can be gained by performing encoded measurements on the QPC-encoded state. We include the encoded measurements as part of the simulation by adding the requisite components to the end of the circuit. The fidelity of this state, which we call the ``encoded fidelity", is a metric that more accurately reflects the quality of the state generation. The purpose of encoded fidelity is not to outperform the logical fidelity, but to show how the encoded information, rather than the physical state, is affected by errors.

In \cref{fig:qpcresult}, we use this metric to show that in certain cases, QPC encodings can protect against distinguishability errors. In this particular example, the resource state is the $X$ eigenstate $|\bar{+}\rangle$, which is protected by increasing the $m$ dimension of the QPC encoding. If the resource state were a $Z$ eigenstate like $|\bar{0}\rangle$, then it would be protected by increasing $n$ instead of $m$. In both cases, the biggest improvement is seen when the QPC encoding reduces to the appropriate repetition code.

\section{Conclusion}
\label{sec:conclusion}
In this work, we have shown that by commuting symmetrization to the end of a circuit, it is possible to significantly reduce the computational cost of a linear optical circuit simulation. We have found that mode assignment lists are a natural framework to handle partial distinguishability because the label degree of freedom can be used as a stand-in for internal modes, which means the photons in the system can have any arbitrary choice of internal states without expanding the Hilbert space. This makes density matrix simulations a practical option for modelling distinguishability errors in linear optical circuits. Density matrices are also a natural way to handle loss and other state-mixing processes without adding extra system modes. We have demonstrated how to implement loss for partially distinguishable photons in a way that avoids any requirement of uniform loss. Many simulations that are otherwise impractical, such as GHZ state generators and QPC-encoding circuits in the presence of arbitrary loss and partial distinguishability, can be performed at much lower computational cost.

The main limitation of using mode assignment lists is, as discussed in \cref{sec:unsym}, that the circuit components must only transform the system's external state (or, in the case of internal mode assignment lists, only its internal state). This prevents a more thorough error model that includes components that act on both the external and internal state, like a beam splitter with frequency dependence. Further work is needed to investigate whether unsymmetrized bases can still provide an advantage in these cases.

We anticipate a broad range of applications for this work. When it comes to linear optical quantum computing, it should be helpful in determining the relative merits of different resource state generation circuits, including boosted circuits. The fact that this method produces a density matrix means we can send it through additional components after the circuit is done, to determine how the errors affect the rest of the resource-generation scheme. Other potential applications include circuit optimization, by maximizing fidelity with respect to the circuit parameters; evaluating error mitigation techniques such as state purification or distillation of indistinguishable photons; and comparing the error tolerance of different encodings. In more general terms, it allows us to determine whether introducing higher resource counts is worth the additional errors being introduced.

There is also the potential for applications beyond linear optical quantum computing. For example, it could be used to study higher-order bunching effects involving larger numbers of photons. It could also apply broadly to quantum communication, metrology, and applications of linear optics supplemented by adaptivity such as learning tasks. Although we exclusively focused on bosons in this paper, an analogous method also applies to fermionic systems if the symmetrization operator is replaced by an anti-symmetrization operator.

\acknowledgments
The authors thank Kamil Br\'{a}dler, Alex Jones, Omkar Srikrishna, Richard Tatham, Joel Wallman, and our colleagues at ORCA Computing for their helpful comments.

\bibliography{apssamp}

\appendix

\onecolumngrid

\section{Gram-Schmidt orthogonalization}
\label{app:gs}

A system of $N$ photons in $M$ external modes is described by a quantum state over a basis that has the information about both the external and internal states of the photons. Given the basis formed by the photons' internal states $\{|\phi_1\rangle,\dots,|\phi_N\rangle\}$, it is possible to perform the simulation by considering the photons effectively distinguishable. Alternatively, each photon's internal state can be decomposed in a different basis. A convenient choice as a basis would be an orthogonal basis. If we consider the basis $\{|\eta_1\rangle,\dots,|\eta_N\rangle\}$, such that $\langle \eta_i| \eta_j \rangle = \delta_{i,j}$, each photon's internal state can be written as 

\begin{equation}
    |\phi_i \rangle = \sum_{j=1}^N \gamma_{i,j} | \eta_j\rangle,
\end{equation}
where $\sum_{j=1}^N |\gamma_{i,j}|^2=1$, leading to $(N-1)$ independent terms and $(N-1)^N$ terms at worse for all the $N$ photons. Alternatively to choosing any random orthogonal basis the number of non-zero coefficients can be reduced to only $(N-1)!$ by using the Gram-Schmidt process~\cite{NielsenBook, Tichy2014}. We defined the orthogonal basis $\{|\tilde\phi_1\rangle,\dots,|\tilde\phi_N\rangle\}$ built via Gram-Schmidt by ascending order such that $\langle \tilde\phi_i| \tilde\phi_j \rangle = \delta_{i,j}$ and 

\begin{equation}
    |\phi_i\rangle = \sum_{j=1}^{i} \tilde\gamma_{i,j} |\tilde\phi_j\rangle
\end{equation}
where the coefficients $\tilde\gamma_{i,j}$ have no physical interpretation as pointed out in \cite{Tichy2015}.

Because photons in the internal state $|\phi_{i}\rangle$ with $i\leq k$ cannot be in the basis state $|\tilde\phi_{j>k}\rangle$, the size of 
$\mathcal{B}_{2Q}$, introduced in \cref{eq:basis-2q}, can be reduced. For example, if we consider the example of a system with two photons in two spatial modes. The internal state is only $|\phi_1\phi_2\rangle$, which is leveraged in the MAL basis written in \cref{eq:MAL_HOM}. If we want to symmetrize and use the $\mathcal{B}_{2Q}$ for instance, we know that using the orthogonalization the possible internal states are $\{|\tilde\phi_1\tilde\phi_1\rangle, |\tilde\phi_1\tilde\phi_2\rangle, |\tilde\phi_2\tilde\phi_1\rangle\}$ and the terms with $|\tilde\phi_2\tilde\phi_2\rangle$ vanishes. This impacts the number of basis states required. We have seen in \cref{eq:2Q_HOM}, we need at most ten terms, but if we remove all the terms arising from $|\tilde\phi_2\tilde\phi_2\rangle$, the basis becomes

\begin{equation}\label{eq:2Q_HOM_GS}
        \mathcal{B}_{2Q}=\{|20\ 00\rangle\rangle,|11\ 00\rangle\rangle,|10\ 10\rangle\rangle,|10\ 01\rangle\rangle,|01\ 10\rangle\rangle,|00\ 20\rangle\rangle,|00\ 11\rangle\rangle\},
\end{equation}
where we have seven terms. Compared to the original ten terms this is a an advantage as the computation is within a smaller space. However, this is not enough to make it competitive with the MAL approach where we only need four terms as shown in \cref{eq:MAL_HOM}. This shows, while processes such as Gram-Schmidt are beneficial, we find that this benefit does not outperform the benefit we gain by using the MAL basis and delaying the symetrization until resolving the interference. We note that using Gram-Schmidt process on the photons' internal state representation has been considered and noted as not a practical option several times \cite{Tichy2015,PhotoniQLAB}. 

\section{Derivation of detection pattern probabilities}
\label{app:det}
\subsection{Detection pattern probabilities in first quantization}
This section contains a derivation of the detection pattern probability in \cref{eq:det-prob}. We will start with the first part of \cref{eq:det-prob}, where the probability of measuring a detection pattern $\bm{d}$ is the sum of probabilities of each corresponding permutation $\pi|\bm{m}_i\rangle$,
\begin{equation}
    P(\bm{d})=\frac{1}{|I_{\bm{m}_i}|}\sum_{\pi\in S_N}\langle\bm{m}_i|\pi^\dagger\rho_\text{ext}\pi|\bm{m}_i\rangle,
\end{equation}
where $|\bm{m}_i\rangle$ can be arranged in any permutation; for example, we can choose the entries in each $|\bm{m}_i\rangle$ to be strictly increasing. The external density matrix is taken from \cref{eq:rho-ext-1q},
\begin{equation}
    \rho_\text{ext}=\frac{1}{|S_N|}\sum_{jk}\frac{\mu_{jk}}{\sqrt{|I_{\bm{m}_j,\bm{\phi}^{(0)}}||I_{\bm{m}_k,\bm{\phi}^{(0)}}|}}\sum_{\sigma,\gamma\in S_N}\sigma|\bm{m}_j\rangle\langle\bm{m}_k|\gamma^\dagger \langle\bm{\phi}^{(0)}|\gamma^\dagger\sigma|\bm{\phi}^{(0)}\rangle,
    \label{eq:rho-ext-app}
\end{equation}
although here we have used different indices. Using the fact that $\pi\rho_{\text{ext}}\pi^\dagger=\rho_{\text{ext}}$, we can write the probability as
\begin{equation}
    P(\bm{d})=\frac{|S_N|}{|I_{\bm{m}_i}|}\langle\bm{m}_i|\rho_{\text{ext}}|\bm{m}_i\rangle.
\end{equation}
Inserting \cref{eq:rho-ext-app} gives
\begin{equation}
        P(\bm{d})=\frac{1}{|I_{\bm{m}_i}|}\sum_{jk}\frac{\mu_{jk}}{\sqrt{|I_{\bm{m}_j,\bm{\phi}^{(0)}}||I_{\bm{m}_k,\bm{\phi}^{(0)}}|}}\sum_{\sigma,\gamma\in S_N}\langle\bm{m}_i|\sigma|\bm{m}_j\rangle\langle\bm{m}_k|\gamma^\dagger|\bm{m}_i\rangle \langle\bm{\phi}^{(0)}|\gamma^\dagger\sigma|\bm{\phi}^{(0)}\rangle
\end{equation}
In order to have nonzero terms, both $|\bm{m}_j\rangle$ and $|\bm{m}_k\rangle$ must have the same set of occupation numbers as $|\bm{m}_i\rangle$, meaning the sum over $j,k$ can be restricted to the submatrix $\mu^{(\bm{d})}$. We can substitute $\sigma\rightarrow\pi_{j\rightarrow i}\sigma$ and $\gamma\rightarrow\pi_{k\rightarrow i}\gamma$, where $\pi_{j\rightarrow i}$ is any chosen permutation that changes $|\bm{m}_j\rangle\rightarrow|\bm{m}_i\rangle$. Then, we can use the fact that
\begin{equation}
    \langle\bm{m}_i|\pi_{j\rightarrow i}\sigma|\bm{m}_j\rangle=
    \begin{cases} 
    1 & \sigma\in I_{\bm{m}_j} \\ 
    0 & \text{otherwise}
    \end{cases}
\end{equation}
and
\begin{equation}
    \langle\bm{m}_k|\gamma^\dagger\pi^\dagger_{k\rightarrow i}|\bm{m}_i\rangle=
    \begin{cases} 
    1 & \gamma\in I_{\bm{m}_k} \\ 
    0 & \text{otherwise}
    \end{cases}
\end{equation}
to eliminate some of the terms in the sum. The remaining terms can be written as a sum over the sets $I_{\bm{m}_j}$ and $I_{\bm{m}_k}$,
\begin{equation}
    P(\bm{d})=\frac{1}{|I_{\bm{m}_i}|}\sum_{jk}\frac{\mu^{(\bm{d})}_{jk}}{\sqrt{|I_{\bm{m}_j,\bm{\phi}^{(0)}}||I_{\bm{m}_k,\bm{\phi}^{(0)}}|}}\sum_{\substack{\sigma\in I_{\bm{m}_j},\\ \gamma\in I_{\bm{m}_k}}} \langle\bm{\phi}^{(0)}|\gamma^\dagger\pi^\dagger_{k\rightarrow i}\pi_{j\rightarrow i}\sigma|\bm{\phi}^{(0)}\rangle.
    \label{eq:Pd-second-last}
\end{equation}
We define
\begin{equation}
    |\bm{\Phi}_j\rangle=\frac{1}{\sqrt{|I_{\bm{m}_j}||I_{\bm{m}_j,\bm{\phi}^{(0)}}|}}\sum_{\sigma\in I_{\bm{m}_j}}\pi_{j\rightarrow i}\sigma|\bm{\phi}^{(0)}\rangle,
    \label{eq:phi-star-2}
\end{equation}
where $\pi_{j\rightarrow i}$ is the permutation that changes $|\bm{m}_j\rangle$ into $|\bm{m}_i\rangle$, which we chose to be the permutation with its external modes in ascending order. This is an alternate definition of \cref{eq:phi-star}, where the internal states are symmetrized within each external mode in order.

To see that \cref{eq:phi-star-2} is equivalent to \cref{eq:phi-star}, consider that since $\sigma\in I_{\bm{m}_j}$, it only permutes photons within the same external mode, never between different external modes. We can use the decomposition $\sum_{\sigma\in I_{\bm{m}_j}}\sigma=\bigotimes_{k=1}^M\sum_{\sigma^{(k)}\in I_{\bm{m}_j}}\pi_{j\rightarrow i}^\dagger\sigma^{(k)}\pi_{j\rightarrow i}$, where $\sigma^{(k)}$ acts on the photons in external mode $k$ and $\pi_{j\rightarrow i}$ re-orders the state to be in order of ascending spatial modes. In other words, the tensor product space is ordered by spatial mode, a permutation is applied within each spatial mode, then the tensor product space is placed back in its original order. We also note that $\pi_{j\rightarrow i}|\phi^{(0)}\rangle=\bigotimes_{k=1}^M|\bm{\phi}^{(0)}_{\mathcal{P}_{k,j}}\rangle$, where $\mathcal{P}_{k,j}$ is the set of photons in spatial mode $k$ for the state $|\bm{m}_j\rangle$. This gives
\begin{equation}
    |\bm{\Phi}_j\rangle=\frac{1}{\sqrt{|I_{\bm{m}_j}||I_{\bm{m}_j,\bm{\phi}^{(0)}}|}}\bigotimes_{k=1}^M\sum_{\sigma^{(k)}\in I_{\bm{m}_j}}\sigma^{(k)}|\bm{\phi}^{(0)}_{\mathcal{P}_{k,j}}\rangle.
\end{equation}
To deal with the normalization constants, we note that $|I_{\bm{m}_j}|=\prod_k|S_{N_k}|$, where $N_k$ is the number of photons in spatial mode $k$, and that $|I_{\bm{m}_{j},\bm{\phi}^{(0)}}|=\prod_k|I_{\bm{\phi}^{(0)}_{\mathcal{P}_{j,k}}}|$. This gives
\begin{equation}
    \begin{split}
        |\bm{\Phi}_j\rangle&=\bigotimes_{k=1}^M\frac{1}{\sqrt{|S_{N_k}||I_{\bm{\phi}^{(0)}_{\mathcal{P}_{j,k}}}|}}\sum_{\sigma^{(k)}\in I_{\bm{m}_j}}\sigma^{(k)}|\bm{\phi}^{(0)}_{\mathcal{P}_{k,j}}\rangle \\
        &=\bigotimes_{k=1}^M\mathcal{S}|\bm{\phi}_{\mathcal{P}_{k,j}^{(0)}}\rangle,
    \end{split}
\end{equation}
which is the same as \cref{eq:phi-star}.

Now that we have shown \cref{eq:phi-star-2} is equivalent to \cref{eq:phi-star}, we can apply \cref{eq:phi-star-2} to \cref{eq:Pd-second-last}, which gives the result
\begin{equation}
    P(\bm{d})=\sum_{jk}\mu_{jk}^{(\bm{d})}\langle\Phi_k|\Phi_j\rangle,
\end{equation}
which matches \cref{eq:det-prob}. Note that $|I_{\bm{m}_{i,j,k}}|$ are all the same for external states with the same detection pattern.

\subsection{Measurement operators}

Alternatively, to obtain the detection pattern probability in \cref{eq:det-prob} we can use measurement operators. To anticipate the normalization condition, it is convenient to use an orthogonal basis for the internal state as described in \cref{eq:ortho_int_basis}. The measurement operator for the spatial mode $k$ has the form
\begin{equation}
    M_{\bm{n}}^{(k)} = \langle\text{vac}|\prod_{i=1}^N  \frac{1}{\sqrt{n_i!}} a_k^{n_i}[\tilde\phi_i],
\end{equation}
where $a_k[\tilde\phi_i]$ is a annihilation operators of a photon with internal state $\tilde\phi_i$ and $\bm{n}$ is a vector indicating how many photons the operator $M_{\bm{n}}$ has for each internal state $\tilde\phi_i$.

The POVM elements are
\begin{equation}
    \Pi_{\bm{n}}^{(k)} = M_{\bm{n}}^{\dagger (k)} M_{\bm{n}}^{(k)} = \prod_{i=1}^N  \frac{1}{\sqrt{n_i!}} a_k^{\dagger n_i}[\tilde\phi_i]|\text{vac}\rangle\langle\text{vac}|\prod_{j=1}^N  \frac{1}{\sqrt{n_j!}} a_k^{n_j}[\tilde\phi_j].
\end{equation}

By definition,
\begin{equation}
    \sum_{\bm{n},|\bm{n}|=N}\Pi_{\bm{n}}^{(k)} = \mathbb{I},
\end{equation}
where we sum over all the vectors $\bm{n}$ giving all possible internal states over $\tilde{\mathcal{B}}_{\text{int}}^{\otimes N}$.

The measurement operator for a detection pattern $\bm{d}$ is
\begin{equation}
    \Pi_{\bm{d}} = \prod_{i=1}^M\left(\sum_{\bm{n},|\bm{n}|=d_i}\Pi_{\bm{n}}^{(i)}\right),
\end{equation}
and the Born rule for a quantum state $\rho$ is

\begin{equation}
    P(\bm{d}) = \text{Tr}(\Pi_{\bm{d}} \rho).
\end{equation}
This leads to the same value as the approach in \cref{eq:det-prob}.

\section{Logical-internal basis}
\label{app:sparrow}
In the logical-internal basis (LI), the state of a system of dual-rail qubits is written as the tensor product of an external state, which is a logical qubit state, and an internal state~\cite{SparrowThesis,Saied2024}. The goal of this section is to put this basis in context and describe its relationship to the other bases we have covered in an attempt to determine when its use is justified.

A Fock state can converted to the logical-internal basis by separating the external and internal state,
\begin{equation}
    |\bm{n}\rangle\rangle\xrightarrow{\text{LI}}|\bm{d}\rangle\rangle\otimes|\bm{\Phi}\rangle,
    \label{eq:LI}
\end{equation}
where $|\bm{d}\rangle\rangle$ is a detection pattern, i.e. a list of occupation numbers for only the external modes, and $|\bm{\Phi}\rangle$ is the sum of all permutations of the ordered internal state $|\bm{\phi}^{(0)}\rangle$ that correspond to the external state $|\bm{d}\rangle\rangle$, as defined in \cref{eq:phi-star}. Note that the logical-internal basis is usually only applied to states where the external state is a logical qubit state; this map is an extension that accepts any Fock state as input and reduces to the standard map for logical states. For example, consider a system with two photons in two external modes. The first few basis states are
\begin{equation}
    \begin{split}
        |20\ 00\rangle\rangle&\xrightarrow{\text{LI}}|20\rangle\rangle|\phi_1\phi_1\rangle \\
        |11\ 00\rangle\rangle&\xrightarrow{\text{LI}}|20\rangle\rangle\left(\frac{|\phi_1\phi_2\rangle+|\phi_2\phi_1\rangle}{\sqrt{2}}\right) \\
        |10\ 10\rangle\rangle&\xrightarrow{\text{LI}}|11\rangle\rangle|\phi_1\phi_1\rangle\\
        |10\ 01\rangle\rangle&\xrightarrow{\text{LI}}|11\rangle\rangle|\phi_1\phi_2\rangle\\
        &\vdots
    \end{split}
\end{equation}
On the right hand side, each Fock state is separated into an external state and an internal state where the internal state is not necessarily symmetric.

In the literature, this basis is sometimes presented as if the Fock state and its logical-internal representation were equivalent. But by converting a Fock state into the logical-internal basis, we break the particle-exchange symmetry. The asymmetry comes from the choice of ordering for the internal states, which is used to create a one-to-one map between Fock states and logical-internal states. As in any unsymmetrized space, photons are effectively distinguishable by their label information. In the logical-internal basis, the labels are chosen such that the external modes are always either ascending or staying the same as the labels increase. The internal states are then ordered accordingly.

Although the logical-internal basis is not an equivalent way to represent a Fock state, it can still be useful. In the literature, it has been used as an intermediate step to trace out internal modes. To show that this use of the logical-internal basis is correct, we begin with a general density matrix $\mu$ in the mode assignment list basis and follow the procedure from \cref{sec:resolve}. Instead of using first quantization, this time we use second quantization:
\begin{equation}
    \begin{split}
        \rho_{2Q}=& \mathcal{S}\mu\mathcal{S}^\dagger \\
        =&\sum_{ij}\mu_{ij}|\bm{n}_i\rangle\rangle\langle\langle\bm{n}_j| \\
        \xrightarrow{\text{LI}}&\sum_{ij}\mu_{ij}|\bm{d}_i\rangle\rangle|\bm{\Phi}_i\rangle\langle\bm{\Phi}_j|\langle\langle\bm{d}_j|.
    \end{split}
\end{equation}
Tracing out the internal modes gives the external density matrix
\begin{equation}
    \begin{split}
        \rho_{\text{ext}}&=\text{Tr}_{\text{int}}(\rho_{2Q}) \\
        &=\sum_{ij}\mu_{ij}|\bm{d}_i\rangle\rangle\langle\langle\bm{d}_j|\langle\bm{\Phi}_j|\bm{\Phi}_i\rangle.
    \end{split}
\end{equation}
The probability of obtaining a detection pattern $\bm{d}$ is
\begin{equation}
    \begin{split}
        P(\bm{d})&=\langle\langle\bm{d}|\rho_{\text{ext}}|\bm{d}\rangle\rangle \\
        &=\sum_{ij}\mu_{ij}^{(\bm{d})}\langle\bm{\Phi}_j|\bm{\Phi}_i\rangle,
    \end{split}
\end{equation}
which is the same result as \cref{eq:det-prob} (derived in \cref{app:det}). So although the internal mode assignment list basis is unsymmetrized, it can still be used as a tool to calculate the correct detection pattern probabilities. However, it is not an equivalent way of writing a second-quantized state. A density matrix in the logical-internal basis would need to be symmetrized in order to be a valid bosonic state.

\section{Circuit connectivity}
\label{app:connectivity}
\subsection{Restricted single-photon spaces}
The main bottleneck in density matrix simulations is the memory required to store large matrices. By working in a smaller, unsymmetrized basis, it is possible to reduce this memory requirement. The space size can be reduced even further using the specific properties of the circuit being simulated. Since the mode assignment list basis is not symmetrized, it is possible to track each photon's location in the circuit using their effective distinguishability. This gives us an intuitive way of restricting the space based on each photon's permitted states. If the system were already symmetrized, then the knowledge of each photon's location in the circuit would be erased.

\begin{figure*}
    \centering
    \includegraphics[trim=0 0 0 0, clip, width=0.8\textwidth]{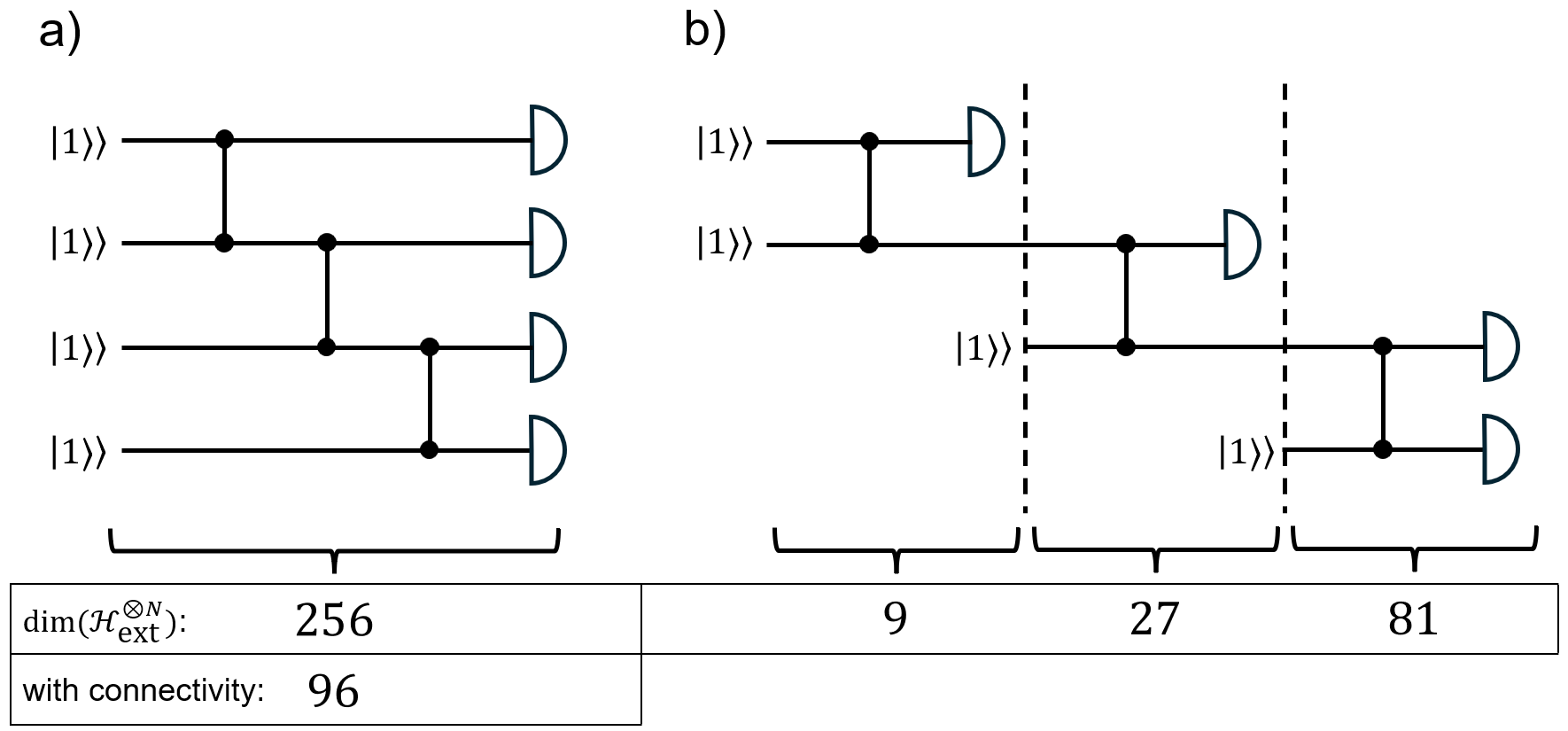}
    \caption{a) An example circuit, chosen only to illustrate the cascading procedure. The number of mode assignment lists $\text{dim}(\mathcal{H}_{\text{ext}}^{\otimes N})=M^N$ is shown below the circuit. This basis size can be reduced using the connectivity of the ciruit; counting from the top of the circuit, $M_1=M_2=4$, $M_3=3$, and $M_4=2$. The resulting space size is significantly smaller. b) The same circuit redrawn in a cascaded layout, which postpones the introduction of certain photons into the circuit until after unconnected external modes have ended in a detector. The number of mode assignment lists is shown for each stage of the simulation, using $\text{dim}(\mathcal{H}_{\text{ext}}^{\otimes N})=(M+1)^N$ to account for the fictitious external mode $0$. By using a cascaded layout, the number of photons and external modes that need to be stored in memory at any one time is reduced. The number of photons is constant throughout the circuit to allow for different combinations of photons arriving at the detectors. This method is made possible by the unsymmetrized representation we are proposing, since the effective distinguishability allows us to know which photons have arrived at a detector. At the end of each stage of the cascaded circuit, the interference is resolved for any photons in the external mode that ends in a detector. In this particular circuit, the space size cannot be further reduced using the circuit connectivity because all the external modes are connected within each stage.}
    \label{fig:cascaded}
\end{figure*}

If we have no prior knowledge of the circuit, then every photon could be located in any of the external modes. This results in the basis $\mathcal{B}_\text{MAL}$, which consists of $M^N$ mode assignment lists. However, each photon can only travel to other external modes when a circuit operation, like a beam splitter or a switch, forms a connection between different modes. Based on these connections, it is possible to restrict each photon's space. Consider a circuit where each of the $i$ photons is allowed to exist in a set of $M_i\leq M$ external modes. In this case, $\mathcal{B}_\text{ext}^{\otimes N}\rightarrow\bigotimes_{i=1}^N\mathcal{B}_{\text{ext},i}$, where $\mathcal{B}_{\text{ext},i}$ is the restricted basis for photon $i$. The size of the $N$-photon space is 
\begin{equation}
    \text{dim}\left(\bigotimes_{i=1}^N\mathcal{B}_{\text{ext},i}\right)=\prod_{i=1}^N M_i\leq M^N.
    \label{eq:size-connectivity}
\end{equation}
For circuits with a low number of connections between modes, this can be a significant reduction. An example of applying this method to a simple circuit is shown in \cref{fig:cascaded}a.

\subsection{Cascaded circuit layouts}
When photons arrive at a detector, they are traced out, as described in \cref{sec:trace-photons}, and then we can post-select on a specific herald pattern. The post-selected state \cref{eq:rho-detected-2} is still in the mode assignment list basis, so it can be sent through more circuit operations before eventually resolving the interference. It is also possible to add in more photons and external modes after this detection process. This technique can be used to simulate larger circuits than would normally fit in the available memory. If there is a subset of photons that can never reach a certain detector, we can wait to add them to the circuit until after the detection has been performed, so the overall state of all photons never needs to be saved. When a circuit is redrawn in such a way that new photons are introduced after others are detected, we will refer to it as being in a \emph{cascaded layout}. An example of a cascaded layout is shown in \cref{fig:cascaded}b.

Using a cascaded circuit layout enables more scalable simulations. Whenever an external mode $m$ ends in a detector, the photons that were in that mode are transferred to the fictitious $0$th mode. Each of those photons has a set of external modes where it is allowed to exist; when one of them arrives at the detector, $m$ is removed from this set and replaced with $0$. If a photon was already allowed in the $0$th mode, then the only effect is to remove $m$. If all of a photon's possible external modes end in detectors, then that photon can only exist in the $0$th mode; this means its set of allowed external modes has size $M=1$, which adds nothing to the total space size \cref{eq:size-connectivity}. This is equivalent to removing the photon from the circuit altogether, making space for more photons to be added in future stages of the circuit.

\section{Loss with partial distinguishability}
\label{app:loss}
\subsection{Derivation of Kraus operators}
We can model the loss of partially distinguishable photons by using a beam splitter to couple the targeted mode to a fictitous loss mode then tracing out the loss mode. Suppose we have a state consisting of $N$ photons in a particular external mode, each of which also has an internal mode,
\begin{equation}
    |\psi\rangle\propto\prod_{i=1}^N a^\dagger(\phi_i)|\text{vac}\rangle\rangle.
    \label{eq:state-before-loss}
\end{equation}
In this expression, $\phi_i$ is the internal state of photon $i$. We ignore the normalization, which depends on the overlaps between the internal modes. We can couple this external mode to the loss mode using a beam splitter operation with transmission probability $\eta$~\cite{Oszmaniec2018}, which gives
\begin{equation}
    \begin{split}
        |\psi'\rangle&\propto\prod_{i=1}^N (\sqrt{\eta}a^\dagger(\phi_i)+\sqrt{1-\eta}a^\dagger_{\text{loss}}(\phi_i))|\text{vac}\rangle\rangle \\
        &=\sum_{n=0}^N\eta^\frac{N-n}{2}(1-\eta)^\frac{n}{2}\sum_{\mathcal{L}_n}\prod_{i\notin\mathcal{L}_n}a^\dagger(\phi_i)\prod_{j\in\mathcal{L}_n}a^\dagger_{\text{loss}}(\phi_j)|\text{vac}\rangle\rangle.
    \end{split}
\end{equation}
In each term in the sum, a certain number $n$ of photons is being lost. When photons are indistinguishable, it does not matter which $n$ photons are lost, but in this case different sets $\mathcal{L}_n$ of $n$ photons give different results, so there is a sum over all possible sets of $n$ photons. Tracing out the loss mode gives
\begin{equation}
    \text{Tr}(|\psi'\rangle\langle\psi'|)\propto\sum_n\eta^{N-n}(1-\eta)^n\sum_{{\mathcal{L}_n},{\mathcal{L}'_n}}\left(\prod_{i\notin{\mathcal{L}_n}}a^\dagger(\phi_i)|\text{vac}\rangle\rangle\langle\langle\text{vac}|\prod_{j\notin\mathcal{L}'_n}a(\phi_j)\right)\left(\langle\langle\text{vac}|\prod_{\ell\in{\mathcal{L}'_n}}a(\phi_\ell)\prod_{k\in{\mathcal{L}_n}}a^\dagger(\phi_k)|\text{vac}\rangle\rangle\right),
    \label{eq:state-after-loss}
\end{equation}
which is the state after undergoing loss.

An equivalent way of applying loss is using a set of Kraus operators. Consider the set $\{K_{\bm{n}}\}$ defined in \cref{eq:kraus},
\begin{equation}
    K_{\bm{n}}=\eta^\frac{N-n}{2}(1-\eta)^\frac{n}{2}\prod_{i}\frac{a^{n_i}(\tilde{\phi}_i)}{\sqrt{n_i!}},
\end{equation}
where $\{\tilde{\phi}_i\}$ is any \emph{orthogonal} set of internal states and $\bm{n}$ is a vector containing the number of lost photons with each orthogonal internal state. The sum of the elements of the loss vector $\bm{n}$ gives the total number of lost photons $n$. To show that this is the correct set of Kraus operators, we will show that applying them to the state gives the correct result, \cref{eq:state-after-loss}. Applying the proposed Kraus operators to the state \cref{eq:state-before-loss} gives
\begin{equation}
    \begin{split}
        \rho&\rightarrow \sum_{\bm{n}} K_{\bm{n}}\rho K^\dagger_{\bm{n}} \\
        &\propto\sum_{\bm{n}}
        \eta^{N-n}(1-\eta)^{n}\prod_{i=1}^{|\bm{n}|}a^{n_i}(\tilde{\phi}_i)
        \prod_{j=1}^N a^\dagger(\phi_j)|\text{vac}\rangle\rangle\langle\langle\text{vac}|\prod_{j'=1}^N a(\phi_{j'}) \prod_{i'=1}^{|\bm{n}|}a^{\dagger n_{i'}}(\tilde{\phi}_{i'}).
    \end{split}
\end{equation}

The creation and annihilation operators can be rearranged using the commutator~\cite{Shchesnovich2022}
\begin{equation}
    [a(\phi_i),a^\dagger(\phi_j)]=\mathbb{I}\langle\phi_i|\phi_j\rangle
\end{equation}
which reduces to the familiar results when the photons are perfectly identical or perfectly distinguishable. Using this commutator with Wick's theorem~\cite{Wick1950} allows us to derive the useful identity
\begin{equation}
    \prod_{i=1}^{|\bm{n}|} a^{n_i}(\tilde{\phi}_i)\prod_{j=1}^N a^\dagger(\phi_j)|\text{vac}\rangle\rangle=\sum_{\mathcal{L}_n}\langle\langle\text{vac}|\prod_{i=1}^{|\bm{n}|}a^{n_i}(\tilde{\phi}_i)\prod_{j\in\mathcal{L}_n} a^\dagger(\phi_j)|\text{vac}\rangle\rangle\prod_{k\notin\mathcal{L}_n} a^\dagger(\phi_k)|\text{vac}\rangle\rangle,
\end{equation}
which gives
\begin{equation}
    \begin{split}
        & \sum_{\bm{n}}
        \eta^{N-n}(1-\eta)^{n}\sum_{\mathcal{L}_n,\mathcal{L}'_n}\left(\prod_{k\notin\mathcal{L}_n} a^\dagger(\phi_k)|\text{vac}\rangle\rangle\langle\langle\text{vac}|\prod_{k'\notin\mathcal{L}'_n} a^\dagger(\phi_{k'})\right) \\
        &\times\langle\langle\text{vac}|\prod_{j'\in\mathcal{L}'_n} a(\phi_{j'})\prod_{i'=1}^{|\bm{n}|}a^{\dagger n_{i'}}(\tilde{\phi}_{i'})|\text{vac}\rangle\rangle\langle\langle\text{vac}|\prod_{i=1}^{|\bm{n}|}a^{n_{i}}(\tilde{\phi}_i)\prod_{j\in\mathcal{L}_n} a^\dagger(\phi_j)|\text{vac}\rangle\rangle \\
        &=\sum_n
        \eta^{N-n}(1-\eta)^{n}\sum_{\mathcal{L}_n,\mathcal{L}'_n}\left(\prod_{k\notin\mathcal{L}_n} a^\dagger(\phi_k)|\text{vac}\rangle\rangle\langle\langle\text{vac}|\prod_{k'\notin\mathcal{L}'_n}a(\phi_{k'})\right)\langle\langle\text{vac}|\prod_{j'\in\mathcal{L}'_n}a(\phi_{j'})\prod_{j\in\mathcal{L}_n}a^\dagger(\phi_j)|\text{vac}\rangle\rangle,
    \end{split}
\end{equation}
the same result as \cref{eq:state-after-loss}. What this shows is that the proposed Kraus operators \cref{eq:kraus} apply the same operation as the beam splitter model. In the last step, we performed the sum over different loss vectors $\bm{n}$, which results in the identity as long as the basis $\{\tilde{\phi}_i\}$ is orthogonal. The particular orthogonal basis being used has no bearing on the result, so we are free to make any choice. One such choice is to orthogonalize the internal states using the Gram-Schmidt process, which is covered in \cref{app:gs}.

\subsection{Derivation of loss map for mode assignment lists}
To derive the loss map \cref{eq:lossmap} we couple the targeted mode to an loss mode $m_\text{loss}$, then place a fictitious detector on the loss mode. There is no heralding at this detector, so the remaining state will be a mixed state of all possible numbers of lost photons. The state before undergoing loss has the form
\begin{equation}
\mu=\sum_{ij}\mu_{ij}|\bm{m}_i)(\bm{m}_j|.
\end{equation}
We will work with one basis state at a time. In each basis state $|\bm{m})$, there is a set $\mathcal{T}$ of photons in the targeted external mode. We can reorder the tensor product to collect these photons together,
\begin{equation}
    |\bm{m})\rightarrow|\bm{m}_{\in\mathcal{T}})\otimes|\bm{m}_{\notin\mathcal{T}}).
\end{equation}
The re-labeled system is the same under symmetrization. The targeted mode is coupled to the loss mode via an imbalanced beam splitter, which implements the transformation
\begin{equation}
    |\bm{m}_{\in\mathcal{T}})\rightarrow\bigotimes_{k\in\mathcal{T}}(\sqrt{\eta}|m_{\in\mathcal{T}})_k+\sqrt{1-\eta}|m_{\text{loss}})_k),
\end{equation}
where the index $k$ indicates which photon is in that mode. This product can be expanded. For each number of photons $n\leq |\mathcal{T}|$, we can choose any subset $\mathcal{L}_{n}\in\mathcal{T}$ of $n$ photons to be lost. Using this notation, the product is
\begin{equation}
    \sum_{n=0}^{|\mathcal{T}|}\eta^{\frac{|\mathcal{T}|-n}{2}}(1-\eta)^\frac{n}{2}\sum_{\mathcal{L}_{n}\in\mathcal{T}}\bigotimes_{k\notin\mathcal{L}_{n}}|m_{\in\mathcal{T}})_k\bigotimes_{k\in\mathcal{L}_{n}}|m_{\text{loss}})_k.
\end{equation}
Using this result, the density matrix after coupling to the loss mode is
\begin{equation}
    \mu=\sum_{ij}\mu_{ij}\sum_{n_i=0}^{|\mathcal{T}_i|}\sum_{n_j=0}^{|\mathcal{T}_j|}\eta^{\frac{|\mathcal{T}_i|-n_i}{2}+\frac{|\mathcal{T}_j|-n_j}{2}}(1-\eta)^\frac{n_i+n_j}{2} \sum_{\substack{\mathcal{L}_{n_i}\in\mathcal{T}_i,\\ \mathcal{L}_{n_j}\in\mathcal{T}_j}}|\bm{m}_{i,\notin\mathcal{L}_{n_i}})|\bm{m}_{i,\in\mathcal{L}_{n_i}})(\bm{m}_{j,\notin\mathcal{L}_{n_j}}|(\bm{m}_{j,\in\mathcal{L}_{n_j}}|.
\end{equation}
The states of all the lost photons in $|\bm{m}_i)$ are combined into the mode assignment list $|\bm{m}_{i,\notin\mathcal{L}_n})$. Next, we can resolve the interference of the photons in the loss mode, symmetrizing and tracing them out as in \cref{eq:rho-detected-2}. The resulting loss map is
\begin{equation}
    \Lambda(\mu)=\sum_{ij}\mu_{ij}\sum_{n=0}^{\text{min}(|\mathcal{T}_i|,|\mathcal{T}_j|)}\eta^{\frac{|\mathcal{T}_i|+|\mathcal{T}_j|}{2}-n}(1-\eta)^n \sum_{\substack{\mathcal{L}_{n_i}\in\mathcal{T}_i,\\ \mathcal{L}_{n_j}\in\mathcal{T}_j}}\langle\bm{\Phi}_{\in\mathcal{L}_{n_j}}|\bm{\Phi}_{\in\mathcal{L}_{n_i}}\rangle|\bm{m}_{i,\notin\mathcal{L}_{n_i}})(\bm{m}_{j,\notin\mathcal{L}_{n_j}}|.
\end{equation}
This applies the same operation as \cref{eq:state-after-loss} but to a density matrix in the mode assignment list basis. To keep track of which photons were traced out, their entries in each mode assignment list can be replaced with $0$.

\end{document}